\documentclass[10pt,journal,final]{IEEEtran}
\usepackage{mathrsfs}
\usepackage{bbding}
\usepackage{graphicx}
\usepackage{float}
\usepackage{booktabs}
\usepackage{hyperref}
\usepackage{cite}

\usepackage{enumitem}
\usepackage{color}
\usepackage{ulem}
\usepackage{psfrag}
\usepackage{subfigure}
\usepackage{amssymb}
\usepackage{amsthm}
\usepackage{setspace}
\usepackage{epsfig}
\usepackage{pifont}
\usepackage{amsmath}
\usepackage{array}
\newcommand{\PreserveBackslash}[1]{\let\temp=\\#1\let\\=\temp}
\newcolumntype{C}[1]{>{\PreserveBackslash\centering}p{#1}}
\newcolumntype{R}[1]{>{\PreserveBackslash\raggedleft}p{#1}}
\newcolumntype{L}[1]{>{\PreserveBackslash\raggedright}p{#1}}
\usepackage{multicol}
\usepackage{multirow}
\usepackage{diagbox}
\usepackage{pifont}
\usepackage{indentfirst}
\usepackage{amsfonts}
\usepackage{algorithm}
\usepackage{algorithmicx}
\usepackage{algpseudocode}
\floatname{algorithm}{Procedure}
\usepackage{fancyhdr}
\usepackage{amscd}
\usepackage[cmyk]{xcolor}
\usepackage{bm}
\usepackage{fancyhdr}
\setlist{itemsep=0pt,parsep=0pt}

\newtheorem{proposition}{Proposition}

\newtheorem{remark}{Remark}

\allowdisplaybreaks[4]

\IEEEoverridecommandlockouts

\usepackage{caption}
\begin{document}
	\title{\huge Deep Learning for Beamforming in Multi-User \\Continuous Aperture Array (CAPA) Systems}
	
	\author{
		\thanks{Jia Guo is with the School of Electronic Engineering and Computer
			Science, Queen Mary University of London, London E1 4NS, U.K. (e-mail:
			jia.guo@qmul.ac.uk).
			
			Yuanwei Liu is with the Department of Electrical and Electronic Engineering,
			The University of Hong Kong, Hong Kong (e-mail: yuanwei@hku.hk).
			
			Hyundong Shin is with the Department of Electronic Engineering, Kyung Hee University, Seoul, South Korea (e-mail:
			hshin@khu.ac.kr).
			
			Arumugam Nallanathan is with the School of Electronic Engineering and Computer
			Science, Queen Mary University of London, London E1 4NS, U.K. (e-mail:
			a.nallanathan@qmul.ac.uk).
			
			A part of this work has been accepted by IEEE Globecom 2024 Workshops \cite{guo2024multi}. If this paper is accepted, we will publicize our codes on GitHub.}
		\IEEEauthorblockN{Jia Guo, \emph{Member, IEEE}, Yuanwei Liu, \emph{Fellow, IEEE}, Hyundong Shin, \emph{Member, IEEE},\\ and Arumugam Nallanathan, \emph{Fellow, IEEE}\vspace{-10mm}}
		
	}
	\maketitle
	\setcounter{page}{1}
	\thispagestyle{empty}
	
	\begin{abstract}
		A DeepCAPA (Deep Learning for Continuous Aperture Array (CAPA)) framework is proposed to learn beamforming in CAPA systems. The beamforming optimization problem is firstly formulated, and it is mathematically proved that the optimal beamforming lies in the subspace spanned by users' conjugate channel responses. Two challenges are encountered when directly applying deep neural networks (DNNs) for solving the formulated problem, i) both the input and output spaces are infinite-dimensional, which are not compatible with DNNs. The finite-dimensional representations of inputs and outputs are derived to address this challenge. ii) A closed-form loss function is unavailable for training the DNN. To tackle this challenge, two additional DNNs are trained to approximate the operations without closed-form expressions for expediting gradient back-propagation.
		To improve learning performance and reduce training complexity, the permutation equivariance properties of the mappings to be learned are mathematically proved. As a further advance, the DNNs are designed as graph neural networks to leverage the properties.
		Numerical results demonstrate that: i) the proposed DeepCAPA framework achieves higher spectral efficiency and lower inference complexity compared to match-filtering and state-of-art Fourier-based discretization method, and ii) DeepCAPA approaches the performance upper bound of optimizing beamforming in the spatially discrete array-based system as the number of antennas in a fixed-sized area tends toward infinity.
		
	\end{abstract}
	
	\section{Introduction}\label{sec:intro}
	
	\IEEEPARstart{T}{o} support higher spectral efficiency (SE) or energy efficiency (EE) in the sixth generation (6G) systems, the massive multi-input-multi-output (MIMO) technique has emerged as a promising solution, where large-scale antenna arrays are employed to improve beam alignment abilities and reduce inter-user interference. Due to the constraint of maintaining half-wavelength spacing between antennas, the deployment of large-scale antenna arrays results in significant fabrication and operational costs.
	
	To address this issue, recent works considered going beyond the classical MIMO composed of spatially discrete arrays (SPD) with half-wavelength spacing. For example, technologies known as holographic MIMO communications \cite{HMIMO-TutorialI-CL2023,HMIMO-6G-WCM2020} or large intelligent surfaces \cite{LIS-TSP2018} have been investigated in recent years.
	By densely packing a large number of low-cost meta-atoms on a small metasurface with distances between the mata-atoms below half-wavelength, a pencil beam with low sidelobe leakage can be formed. Hence, high system performance such as SE can be achieved with a low cost, size, weight, and low power consumption hardware architecture \cite{HMIMO-BF-TWC2022}.
	
	The asymptotic form of holographic MIMO is continuous aperture arrays (CAPA), which leverages programmable metamaterials for approximately realizing a continuous microwave aperture \cite{HMIMO-6G-WCM2020}. CAPA is expected to approach the ultimate capacity limit of wireless
	channels \cite{zhao2024continuous}. Recent works on CAPA communications mainly focus on performance analysis \cite{ouyang2024diversity,zhao2024continuous,wan2023mutual}, channel estimation \cite{HMIMO-TutorialI-CL2023} and system modeling \cite{Chl_model_HMIMO-TWC2022,HMIMO-TutorialI-CL2023,CAPA-SEOpt-TWC2024}. In the CAPA systems, the current distribution is a key factor affecting the system performance \cite{Comm_LIS_2020}, which is analogous to beamforming for SPD systems. In what follows, we refer to current distributions in CAPA systems as beamforming, whose optimization is challenging. This is because functional optimization problems need to be solved to find the optimal solutions, where traditional convex optimization tools are not applicable.
	Existing works strive to tackle this challenge under specific scenarios or assumptions, such as the single- and two-user scenarios \cite{zhao2024continuous}, and the assumption that the optimal beamforming can be represented by the Fourier basis functions \cite{sanguinetti2022wavenumber}. How to design beamforming in general scenarios is under-investigated.
	
	\vspace{-1mm}\subsection{Related Works}
	\subsubsection{Optimizing Beamforming for CAPA}
	
	Recent works made efforts to solve the beamforming optimization problems numerically.
	In \cite{zhao2024continuous}, a closed-form beamforming solution was derived under single- and two-user cases, by first finding the optimal uplink decoder and then resorting to uplink-downlink duality. In \cite{sanguinetti2022wavenumber}, a line-of-sight (LoS) single-user multi-stream CAPA communication system was considered, where the beamforming solutions can be generated by Fourier basis functions, and the powers of the functions were optimized with water-filling to maximize SE.
	
	When it comes to multi-user scenarios with more than two users, 
	The beamforming optimization problem was solved numerically in \cite{zhang2023pattern} and \cite{CAPA-SEOpt-TWC2024}, where the key idea is to use Fourier expansion to project the continuous spatial response onto an orthogonal basis space, such that the original functional optimization problem is reduced to a parameter optimization problem aiming at optimizing the projection lengths on finite orthogonal bases. Since the number of basis functions grows exponentially with the CAPA aperture size and signal frequency, this approach incurs prohibitive computational complexity for solving high-dimensional problems \cite{wang2024beamforming}. A sub-optimal method to the functional optimization problem was compared in \cite{sanguinetti2022wavenumber} as a performance baseline, where the CAPA was divided into multiple patches such that the problem can be reduced to one in the SPD system. This discretization yields significant performance loss.
	
	\subsubsection{Learning Wireless Policies with Deep Neural Networks}
	Existing works designed deep neural networks (DNNs) to learn wireless policies (i.e., the mapping from the environmental parameters such as channel coefficients to decisions such as precoding) from problems in SPD systems, including fully connected neural networks (FNNs) \cite{FNN_mUE_DigPre_TVT2021_Kong, FNN_mUE_DigPre_WCL_2020_Kim}, convolutional neural networks (CNNs) \cite{CNN_sUE_HyPre_TCOM2022_Liu} and graph neural networks (GNNs) \cite{GNN-PC-CellFree-TWC2024, GNN-PA-CellFree-TWC2024,GAT-RIS-Precoding-TVT2024,GAT-MU-MISO-Precoding-TWC2024}. The deep learning methods are superior to traditional optimization-based methods in low inference time \cite{FNN_mUE_DigPre_WCL_2020_Kim, LSJ_MultiDim_GNN_2022}, ability of jointly optimizing with channel acquisition \cite{FNN_mUE_DigPre_TVT2021_Kong,GNN-Pilot-precoding-ICC2024}, or improving the robustness to channel estimation errors \cite{CNN_sUE_HyPre_TCOM2022_Liu}. 
	
	Compared to FNNs and CNNs, GNNs show superiority in generalizability to unseen graph sizes \cite{GNN-PA-CellFree-TWC2024,GAT-MU-MISO-Precoding-TWC2024}, scalability to large scale systems \cite{GNN-PC-CellFree-TWC2024} and better learning performance \cite{GJ_TWC_GNN}. 
	It has been noticed that the advantages of GNNs steam form harnessing the permutation equivariance (PE) properties that widely exist in wireless policies \cite{GJ_TWC_GNN, LSJ_MultiDim_GNN_2022}. 
	The PE property satisfied by a GNN is affected by graph modeling that determines which vertices and edges are of the same type. Many existing works of learning wireless policies with GNNs model graphs directly from the wireless networks \cite{GNN-PA-CellFree-TWC2024,PC-CA-TWC-2024,GNN-Pilot-precoding-ICC2024}, e.g., the base stations (BSs) and the users are modeled as two types of vertices and the links between them are modeled as edges. Such a heuristic graph modeling may cause a property mismatch issue. Specifically, the input-output relationship of a GNN does not satisfy the property of the wireless policy to be learned, which leads to degradation of learning performance \cite{GJ_TWC_GNN}.
	To avoid such a property mismatch issue,  a scheme of systematically modeling graphs from optimization problems was proposed in \cite{LSJ_MultiDim_GNN_2022}.
	
	\vspace{-2mm}
	\subsection{Motivations and Contributions}
	Motivated by the abilities of DNNs to learn unknown relationships from massive training data, in this paper, we propose DNNs to learn beamforming for CAPA systems.
	As far as the authors know, there are no works addressing learning beamforming (or any other wireless policies) in CAPA systems.
	Directly applying DNNs to learn the beamforming policy (i.e., the relationship from channel functions to beamforming solutions) encounters two challenges as follows. 
	\begin{enumerate}
		\item Both the channel responses and the beamforming solutions are functions that can be regarded as vectors with infinite dimensions, which cannot be directly inputted into or outputted from DNNs.
		\item To avoid generating labeled training data that is time-consuming, the DNNs are trained in an unsupervised manner. By using the objective function of the optimization problem as the training loss, the integrals in the objective function without closed-form expressions hinder back-propagation.
	\end{enumerate}
	To address these two challenges, we propose a deep learning framework called DeepCAPA. Our main contributions are summarized as follows. 
	\begin{itemize}
		\item We identify and investigate beamforming optimization in multi-user CAPA systems by using deep learning. We first formulate a functional optimization problem, and mathematically prove that the optimal beamforming lies in the subspace spanned by users' conjugate channel responses. Such a property has not been revealed in the literature of optimizing beamforming in CAPA systems, and it helps tackle the first challenge.
		\item We propose the DeepCAPA framework to learn the beamforming policies for multi-user CAPA systems. To tackle the first challenge, we find vector representations for channel responses and beamforming. Regarding the second challenge, we train  two additional DNNs to approximate the integrals in the objective functions and constraints for expediting back-propagation.
		\item We mathematically prove the PE properties inherent in the mappings to be learned, and design the DNNs as GNNs that are incorporated with these properties to improve learning performance. To ensure that proper PE properties are embedded into the GNNs for efficient learning, we construct the graphs from the properties instead of heuristically from the CAPA-based system.
		\item The results show that the proposed framework for learning a SE-maximal beamforming achieves the upper bound of SE in the SPD system as the number of antennas in a fixed-sized area approaches infinity. The proposed framework also outperforms the existing methods by providing significantly higher SE than match-filtering and shorter inference time than the Fourier-based method in \cite{zhang2023pattern}, which enables real-time implementation.
	\end{itemize}
	
	While a part of this work has been published in \cite{guo2024multi}, the contents in this journal version are substantially extended, including revealing the permutation property mismatch issue, proposing an alternative training way for better learning performance, and comparing the performance with more numerical algorithms under various settings in the simulations to better understand when the proposed framework is with larger performance gain.

	The rest of the paper is organized as follows. In section \ref{sec: system model}, we introduce the system model and the beamforming policy. In section \ref{sec:learn-bf}, we show how to design the DeepCAPA framework for learning beamforming. In section \ref{sec:gnn}, we show how to learn the mappings in the framework over graphs with GNNs. In section \ref{sec:results}, we provide numerical results.
	In section \ref{sec:conclusions}, we provide the conclusion remarks.

	\vspace{-1mm}
	\section{System Model and Beamforming Policy}\label{sec: system model}\vspace{-1mm}
	Consider a downlink system where a BS transmits to $K$ users. Both the BS and each user are equipped with a CAPA. The aperture spaces of the CAPAs at the BS and the $k$-th user are denoted as $\mathcal{A}$ and $\mathcal{A}_k\subseteq {\mathbb R}^{3\times 1}$, respectively, and the CAPA of the $k$-th user is centered at $\mathbf{s}_k \in {\mathbb R}^{3\times 1}$.
	
	\subsection{Signal Model}
	The transmitted signal at point $\mathbf{r}\in\mathcal{A}$ can be expressed as $x(\mathbf{r})=\sum_{k=1}^K \mathsf{V}_k(\mathbf{r})s_k$, where $s_k$ is the symbol to be transmitted to the $k$-th user with ${\mathbb E}\{|s_{k}|^2\}=1$, and $\mathsf{V}_k(\mathbf{r}) (\mathbf{r}\in\mathcal{A})$ is the beamforming to convey $s_k$. 
	
	The observation of the $k$-th user at point $\mathbf{s}\in\mathcal{A}_k$ can be written as,
	\begin{eqnarray}\label{eq:receive-sig}
		\mathsf{Y}_k(\mathbf{s}) \!\!&\!\!=\!\!&\!\! \textstyle\int_{\mathcal{A}} \mathsf{H}(\mathbf{r,s})x(\mathbf{r})d\mathbf{r} + \mathsf{N}_k(\mathbf{s}) \notag\\
		\!\!&\!\!=\!\!&\!\! \textstyle\int_{\mathcal{A}} \mathsf{H}(\mathbf{r,s})\Big(\textstyle\sum_{j=1}^K \mathsf{V}_j(\mathbf{r})s_j\Big) d\mathbf{r} + \mathsf{N}_k(\mathbf{s})\notag\\
		\!\!&\!\!\overset{(a)}{\approx}\!\!&\!\!
		\underbrace{\textstyle\int_{\mathcal{A}} \mathsf{H}_k(\mathbf{r}) \mathsf{V}_k(\mathbf{r})s_k d\mathbf{r}}_{(b)} + \notag\\
		&& \underbrace{\textstyle\sum_{j=1,j\neq k}^K\int_{\mathcal{A}} \mathsf{H}_k(\mathbf{r}) \mathsf{V}_j(\mathbf{r})s_j d\mathbf{r}}_{(c)} + \mathsf{N}_k(\mathbf{s}),
	\end{eqnarray}
	where $\mathsf{H}(\mathbf{r,s})$ is the space channel response from $\mathbf{r}$ to $\mathbf{s}$, and  $(a)$ comes from splitting the observation by the received signal (i.e., term $(b)$), interference from other users (i.e., term $(c)$) and thermal noise $\mathsf{N}_k(\mathbf{s})$ satisfying ${\mathbb E}\{\mathsf{N}_k(\mathbf{s})\mathsf{N}_k^*(\mathbf{s}')\}=\sigma_k^2\delta(\mathbf{s}-\mathbf{s}')$, $(\cdot)^*$ denotes conjugate of a complex value. The approximation in $(a)$ comes from the assumption that the aperture size of each user is smaller enough compared to the propagation distance and the BS aperture size, such that the channel response to each point on the $k$-th user's aperture is almost the same as the channel response to the center of the $k$-th user's aperture $\mathbf{s}_k$, i.e., $\mathsf{H}(\mathbf{r,s})\approx\mathsf{H}(\mathbf{r,s}_k)\triangleq \mathsf{H}_k(\mathbf{r})$.
	
	With \eqref{eq:receive-sig}, the signal-to-interference-and-noise ratio of the $k$-th user can be approximated as \cite{zhao2024continuous},
	\begin{eqnarray}\label{eq:sinr}
		\gamma_k 
		\approx
		\frac{|\mathcal{A}_k|\cdot|\int_{\mathcal{A}} \mathsf{H}_k(\mathbf{r}) \mathsf{V}_k(\mathbf{r}) d\mathbf{r}|^2}{\sum_{j=1,j\neq k}^K |\mathcal{A}_j|\cdot|\int_{\mathcal{A}} \mathsf{H}_k(\mathbf{r}) \mathsf{V}_j(\mathbf{r}) d\mathbf{r}|^2 + \sigma_k^2},
	\end{eqnarray}
	where $|\mathcal{A}_k|$ is the aperture size of the $k$-th user. The approximation still comes from the assumption that the aperture size of each user is smaller enough compared to the propagation distance and the BS aperture size. 
	
\begin{figure}[!htb]
		\centering
		\includegraphics[width=\linewidth]{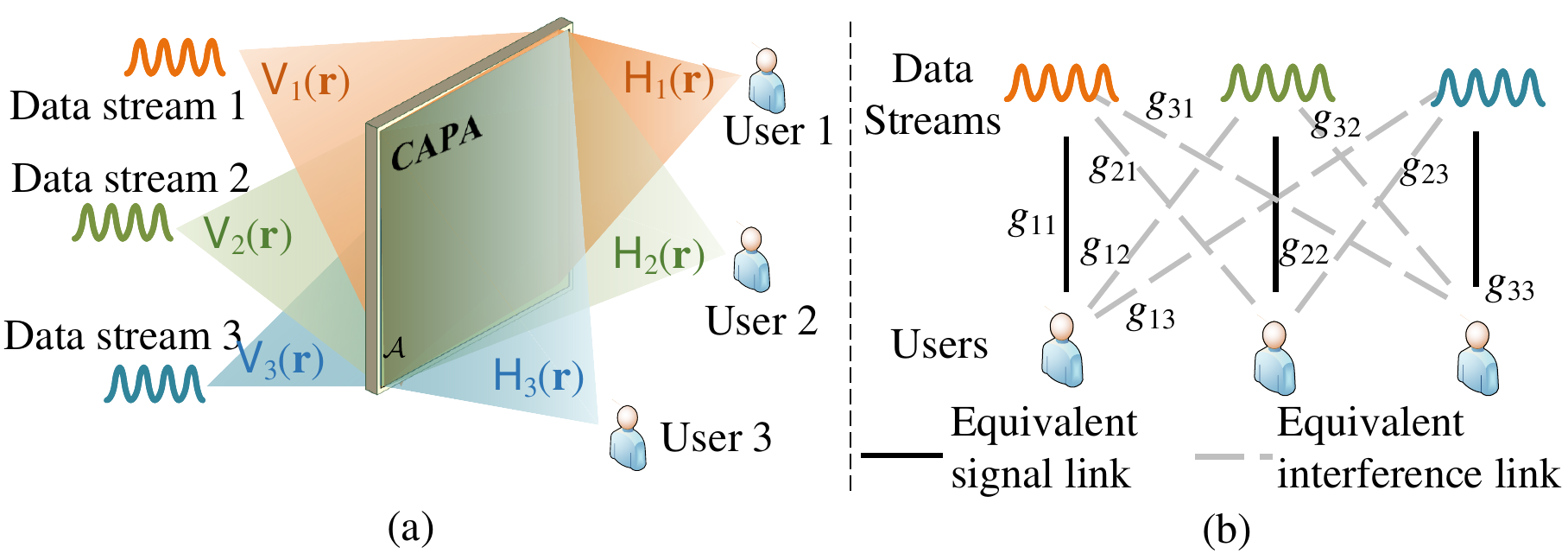}
		\vspace{-5mm}
		\caption{(a) An example CAPA transmission system with three data streams transmitted to three users. (b) The equivalent system.}
		\vspace{-3mm}
		\label{fig:capa}
	\end{figure}

	The CAPA communication system can be regarded as an equivalent system where $K$ data streams are transmitted to $K$ users. The $k$-th data stream is transmitted to the $k$-th user through an equivalent signal link with channel gain of $g_{kk}=\int_{\mathcal{A}}\mathsf{H}_k(\mathbf{r})\mathsf{V}_k(\mathbf{r})d\mathbf{r}$, and is interfered by other transmitted data streams through equivalent interference links with channel gains of  $g_{kj}=\int_{\mathcal{A}}\mathsf{H}_k(\mathbf{r})\mathsf{V}_j(\mathbf{r})d\mathbf{r}, j\neq k$. An example CAPA communication system and the equivalent system is illustrated in Fig. \ref{fig:capa}.
	
	\subsection{Channel Model}
	To simplify our analysis, we consider that there are only LoS channels between the BS and the users. Then, $\mathsf{H}_k(\mathbf{r})$ can be modeled as follows \cite{zhao2024continuous},
	\begin{eqnarray}\label{eq:chn-model}
		\mathsf{H}_k(\mathbf{r}) \!\!&\!\!=\!\!&\!\! \sqrt{\frac{\mathbf{e}_r^T(\mathbf{s}_k-\mathbf{r})}{\|\mathbf{r}-\mathbf{s}_k\|}}\cdot\notag\\
		&&\frac{jk_0\eta e ^{-jk_0\|\mathbf{r\!-\!s}_k\|}}{4\pi\|\mathbf{r}-\mathbf{s}_k\|}\Bigg(\!1\!+\!\frac{j/k_0}{\|\mathbf{r}-\mathbf{s}_k\|}\!-\!\frac{1/k_0^2}{\|\mathbf{r}-\mathbf{s}_k\|^2}\!\Bigg)\!,
	\end{eqnarray}
	where  ${\mathbf{e}_r}\in{\mathbb{R}}^{3\times1}$ is the normal vector of the CAPA at the BS, $\eta=120\pi$ is the impedance of free space, $k_0=\frac{2\pi}{\lambda}$ with $\lambda$ being the wavelength denotes the wavenumber, and $(\cdot)^T$ denotes matrix transpose.
	
	\subsection{Beamforming Optimization Problem and Policy}
	
	With \eqref{eq:sinr}, we can formulate a problem to optimize beamforming for all the users. Taking maximizing SE as an example, the problem can be formulated as,
	\begin{subequations}
		\begin{align}
			{\bf P1}:
			\max_{\mathsf{V}_k(\mathbf{r}),k=1,\cdots,K}~~ & \textstyle\sum_{k=1}^K \log_2 \left(1+\gamma_k \right) \label{eq:bb-object} \\
			{\mathrm{s.t.}} ~~
			&  \textstyle\sum_{k=1}^K\int_{\mathcal{A}}|\mathsf{V}_k(\mathbf{r})|^2 d\mathbf{r} =  P_{\max},\label{eq:bb-constraint}
		\end{align}
	\end{subequations}
	where $P_{\max}$ is the maximal transmit power of the BS.
	
	The beamforming policy is referred to as the mapping from the space channel responses to the beamforming solutions, which is denoted as a function $\mathsf{V}=F_{\mathsf{BF}}(\mathsf{H})$, where $\mathsf{V}(\cdot)=[\mathsf{V}_1(\cdot),\cdots,\mathsf{V}_K(\cdot)]$ and $\mathsf{H}(\cdot)=[\mathsf{H}_1(\cdot),\cdots,\mathsf{H}_K(\cdot)]$ are multi-variate functions. We do not express the beamforming policy as $\mathsf{V}(\mathbf{r})=F_{\mathsf{BF}}(\mathsf{H}(\mathbf{r}))$, which may cause misunderstanding that the beamforming at point $\mathbf{r}$ is only related to the channel response at this point.

	\begin{proposition}\label{prop1}
		The optimal beamforming is in a function subspace spanned by $\mathsf{H}_1^*(\mathbf{r}),\cdots,\mathsf{H}_K^*(\mathbf{r})$, i.e., $\mathsf{V}_k(\mathbf{r})$ can be expressed as $\mathsf{V}_k(\mathbf{r})=\sum_{j=1}^K b_{jk} \mathsf{H}_j^*(\mathbf{r})$, where $b_{jk}$ is a scalar linear combination coefficient.
		\begin{IEEEproof}
			See Appendix \ref{proof:prop1}.
		\end{IEEEproof}
	\end{proposition}
	
	\section{Learning to Optimize Beamforming}\label{sec:learn-bf}
	
	In this section, we propose a deep learning-based method to learn the beamforming policy $\mathsf{V}=F_{\mathsf{BF}}(\mathsf{H})$. We first show two challenges encountered when directly applying conventional deep learning framework as in \cite{GJ_TWC_GNN}, and then illustrate how to tackle these challenges with our proposed framework.
	
	\begin{figure*}[!htb]
		\centering
		\includegraphics[width=.9\linewidth]{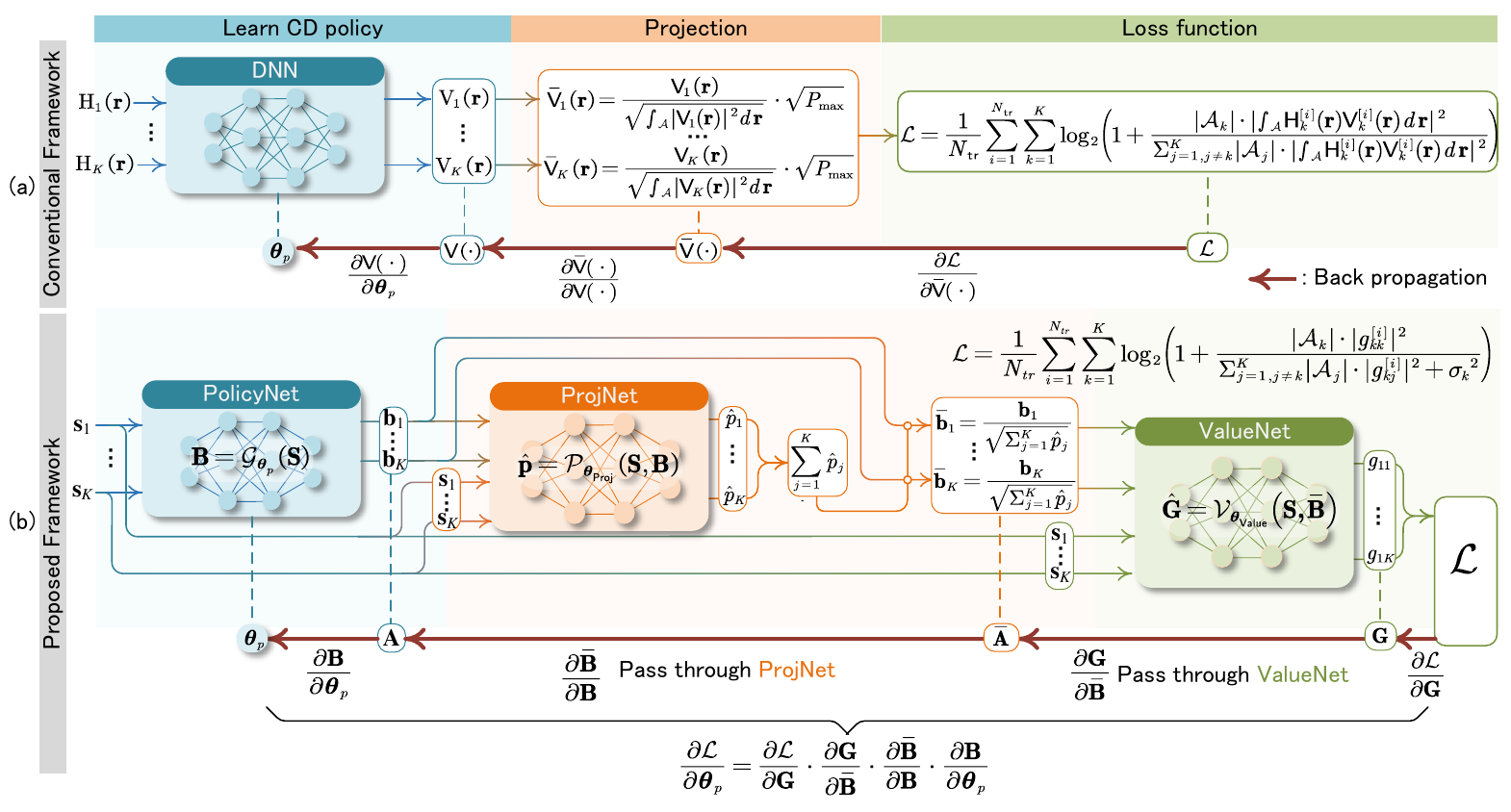}
		\vspace{-3mm}
		\caption{Illustration of (a) the conventional deep learning framework and (b) the proposed DeepCAPA.}
		\vspace{-6mm}
		\label{fig:dl-arch}
	\end{figure*}
	
	\subsection{\!Challenges of Applying Conventional Deep Learning Framework}
	Intuitively, a DNN can be applied to learn the beamforming policy, where the input and output of the DNN are respectively  $\mathsf{H}(\cdot)$ and $\mathsf{V}(\cdot)$, as shown in Fig. \ref{fig:dl-arch}(a). Since the learned beamforming may not satisfy the power constraint in \eqref{eq:bb-constraint}, we normalize the learned beamforming as follows such that it can be projected to the region satisfying the constraint,
	\begin{equation}\label{eq:act-func}
		\bar{\mathsf{V}}_k(\mathbf{r})=\mathsf{V}_k(\mathbf{r})\sqrt{P_{\max}}\Big/\sqrt{\textstyle\sum_{j=1}^K \textstyle\int_{\mathcal{A}}|\mathsf{V}_j(\mathbf{r})|^2 d\mathbf{r}}.
	\end{equation}
	
	To avoid generating labeled training data that is time-intensive,
	the DNN is trained in an unsupervised manner as in \cite{GJ_TWC_GNN}. The loss function can be the objective function in \eqref{eq:bb-object} averaged over all the training samples, i.e., 
	
	{\small
		\begin{equation}\label{eq:loss-func}
			\mathcal{L}\!=\!\frac{1}{N_{\mathsf{tr}}}\!\sum_{i=1}^{N_{\mathsf{tr}}}\!\sum_{k=1}^K\!\log_2\!\Bigg(\!1+\frac{|\mathcal{A}_k|\cdot|\int_{\mathcal{A}} \mathsf{H}_k^{[i]}(\mathbf{r}) \bar{\mathsf{V}}_k^{[i]}(\mathbf{r}) d\mathbf{r}|^2}{\sum\limits_{j=1,j\neq k}^K |\mathcal{A}_j|\!\cdot\!|\int_{\mathcal{A}} \mathsf{H}_k^{[i]}(\mathbf{r}) \bar{\mathsf{V}}_j^{[i]}(\mathbf{r}) d\mathbf{r}|^2 \!+\! \sigma_k^2}\Bigg),
	\end{equation}}
	
	\noindent where $N_{\mathsf{tr}}$ denotes the number of training samples, and the superscript $[i]$ denotes the $i$-th sample.
	
	However, such a design encounters two challenges,
	\begin{itemize}[leftmargin=0.4cm]
		\item \emph{Challenge 1: $\mathsf{H}(\cdot)$ and $\mathsf{V}(\cdot)$ are functions that can be seen as vectors with infinite dimensions, whereas the inputs and outputs of DNNs should be with finite dimensions}. 
		\item \emph{Challenge 2: Both \eqref{eq:act-func} and \eqref{eq:loss-func} include integrals that are without closed-form expressions}. This hinders computing the gradients (specifically, $\frac{\partial \mathcal{L}}{\partial \bar{\mathsf{V}}^{[i]}_k(\cdot)}$ and $\frac{\partial\bar{\mathsf{V}}^{[i]}_k(\cdot)}{\partial \mathsf{V}^{[i]}_k(\cdot)}$ in Fig. \ref{fig:dl-arch}(a)) for training the DNN with back-propagation.
	\end{itemize}
	
	\subsection{The Proposed Deep Learning Framework}\label{sec:proposed-framework}
	To tackle these two challenges, we propose a deep learning framework to learn beamforming policies, which is called Deep Learning for CAPA (DeepCAPA). The core idea for tackling Challenge 1 is to find finite-dimensional representations of the functions that can be inputted into and outputted from DNNs. The core idea for tackling Challenge 2 is to learn the integrals with DNNs. Since the input-output relationships of DNNs only include basic operations such as matrix addition and multiplication, the relationships are with closed-form expressions, and their gradients can be computed for learning the policy.
	
	From the analysis above, the framework of learning beamforming should include three DNNs, (i) a PolicyNet to learn a finite-dimensional representation of $\mathsf{V}(\cdot)$ from a finite-dimensional representation of $\mathsf{H}(\cdot)$, (ii) a ProjNet to learn the integrals in \eqref{eq:act-func}, and (iii) a ValueNet to learn the integrals in \eqref{eq:loss-func}. The overall framework is shown in Fig. \ref{fig:dl-arch}(b). In the following, we will introduce the neural networks in detail.
	
	\subsubsection{PolicyNet}\label{sec:policynet}
	The PolicyNet is designed to learn the mapping from $\mathsf{H}(\cdot)$ to $\mathsf{V}(\cdot)$. To this end, we need to tackle \emph{Challenge 1} such that only finite-dimensional vectors are inputted into and outputted from the PolicyNet. One way is to find the ``bases'' of these two functions.
	
	\textbf{Finding finite-dimensional representation of $\mathsf{H}(\cdot)$:} It can be seen from the channel model \eqref{eq:chn-model} that $\mathsf{H}_k(\mathbf{r})$ only depends on  $\mathbf{e}_r, \eta, k_0$ and the locations of the $k$-th user's aperture, i.e., $\mathbf{s}_k$. Since $\mathbf{e}_r, \eta, k_0$ can be seen as constants in the CAPA communication system, $\mathsf{H}_k(\mathbf{r})$ only varies with $\mathbf{s}_k$, and $\mathsf{H}_k(\mathbf{r})$ is known if $\mathbf{s}_k$ is known. Hence, $\mathbf{s}_k\in{\mathbb R}^{3\times 1}$ can be used as a finite-dimensional representation of $\mathsf{H}_k(\mathbf{r})$.
	
	\textbf{Finding finite-dimensional representation of $\mathsf{V}(\cdot)$:} To this end, we resort to Proposition \ref{prop1} that the optimal beamforming is a linear combination of users' channel functions. 
	This indicates that once $\mathsf{H}_1(\mathbf{r}),\cdots,\mathsf{H}_K(\mathbf{r})$ are known, we can obtain $\mathsf{V}_k(\mathbf{r})$ with $\mathbf{b}_k\triangleq [b_{1k},\cdots,b_{Nk}]^T$, which is called the \emph{linear combination vector} of the $k$-th user. Hence, $\mathbf{b}_k$ can be used as a finite-dimensional representation of $\mathsf{V}_k(\mathbf{r})$.
	
	After finding the finite-dimensional representations of  $\mathsf{H}(\cdot)$ and $\mathsf{V}(\cdot)$, we can learn the beamforming policy by learning $\mathbf{B}\triangleq[\mathbf{b}_1,\cdots,\mathbf{b}_K]\in {\mathbb C}^{K\times K}$ from $\mathbf{S}\triangleq[\mathbf{s}_1,\cdots,\mathbf{s}_K]^T\in {\mathbb C}^{K\times 3}$. The policy to be learned is denoted as $\mathbf{B}=F_{\mathsf{P}}(\mathbf{S})$.
	
	
	\begin{proposition}\label{prop2}
		The beamforming policy satisfies the following permutation property,
		\begin{eqnarray}\label{eq:perm-policy}
			{\bm\Pi}^T\mathbf{B}{\bm\Pi} = F_{\mathsf{P}}({\bm\Pi}^T\mathbf{S}),
		\end{eqnarray}
		where ${\bm \Pi}$ is an arbitrary permutation matrix\footnote{An example of permutation matrix is $\bm\Pi=\textstyle\Big[\begin{smallmatrix}
				0&1&0\\1&0&0\\0&0&1
			\end{smallmatrix}\Big]$, which changes the order of elements in a vector $\mathbf{x}=[x_1,x_2,x_3]^T$ to $\bm\Pi^T\mathbf{x}=[x_2,x_1,x_3]^T.$}
		\begin{IEEEproof}
			See Appendix \ref{proof:prop1}.
		\end{IEEEproof}
	\end{proposition}
	The permutation property in \eqref{eq:perm-policy} can be understood as follows. As can be seen from Fig. \ref{fig:capa}, when the order of users is permuted by a matrix $\bm \Pi$, the $k$-th user is permuted to the $\pi(k)$-th user \footnote{$\pi(\cdot)$ is an operator that changes the order of indices of users, such that $\bm \Pi^T[1,\cdots,K]^T=[\pi(1),\cdots,\pi(K)]^T$.}. Only when the order of data streams is permuted by the same matrix, the serving relationship of the CAPA system remain unchanged, i.e., the $\pi(k)$-th data stream that is permuted from the $k$-th data stream is still transmitted to the $\pi(k)$-th user. Then, the summation of data rate in \eqref{eq:bb-object} and the beamforming policy remain unchanged. Hence, the policy satisfies a \emph{dependent permutation property} \cite{LSJ_MultiDim_GNN_2022}, where the permutations of the rows and columns in $\mathbf{B}$ (respectively correspond to the permutations of users and data streams) depend on the same permutation matrix.
	
	
	\textbf{Neural Network Structure:} The input and output of the PolicyNet are respectively $\mathbf{S}$ and $\mathbf{B}$. The PolicyNet can be a FNN, or a GNN that leverages the permutation property in \eqref{eq:perm-policy} for better learning performance, as to be detailed in section \ref{sec:gnn}. The input-output relationship of the PolicyNet is denoted as,
	\begin{equation}\label{eq:PolicyNet}
		\mathbf{B}=\mathcal{G}_{\bm\theta_p}(\mathbf{S}),
	\end{equation}  
	where $\bm\theta_p$ denotes all the free parameters (e.g., the weight matrices) in the PolicyNet.
	
	The training of $\bm\theta_p$ depends on the trained ProjNet and ValueNet. Hence, we first introduce these two neural networks, and then introduce how to train the PolicyNet in section \ref{sec:training}.
	
	\subsubsection{ProjNet}
	The learned finite-dimensional representations of beamforming (i.e., the linear combination vectors) need to be normalized as in \eqref{eq:act-func} such that it is projected to the feasible region of problem \textbf{P1}. To tackle Challenge 2 that the integral in \eqref{eq:act-func} (i.e., the power consumed by $\mathsf{V}_k(\mathbf{r})$) hinders back-propagation, we design the ProjNet to learn power consumption as follows. 
	
	The ProjNet aims to learn the mapping from the finite-dimensional representations of $\mathsf{V}_k(\mathbf{r}),k=1,\cdots,K$ to the power consumption for each user. As have been analyzed in section \ref{sec:policynet}, $\mathsf{H}_k(\mathbf{r})$ is known if $\mathbf{s}_k$ is known, and after $\mathsf{H}_k(\mathbf{r}),k=1,\cdots,K$ are known, we can obtain $\mathsf{V}_k(\mathbf{r}),k=1,\cdots,K$ with $\mathbf{B}$. Then, $\mathbf{S}$ and $\mathbf{B}$ can be the representation of $\mathsf{V}_k(\mathbf{r}),k=1,\cdots,K$, and the mapping to be learned by the ProjNet is denoted as $\mathbf{p}=F_{\mathsf{Proj}}(\mathbf{S,B})$, where $\mathbf{p}=[p_1,\cdots,p_K]^T$ is the vector of power consumption of the users. 
	
	With the same way of analyzing the permutation property of the beamforming policy in \eqref{eq:perm-policy}, we can obtain the property satisfied by the function $F_{\mathsf{Proj}}(\cdot)$ in the following proposition,
	\begin{proposition}\label{prop3}
		$F_{\mathsf{Proj}}(\cdot)$ satisfies the following permutation property,
		\begin{equation}\label{eq:perm-proj}
			{\bm\Pi}_2^T\mathbf{p}=F_{\mathsf{Proj}}({\bm\Pi}_1^T\mathbf{S}, {\bm\Pi}_1^T\mathbf{B}{\bm\Pi}_2),
		\end{equation}
		where $\bm\Pi_1$ and $\bm\Pi_2$ are permutation matrices.
		\begin{IEEEproof}
			See Appendix \ref{proof:prop3}.
		\end{IEEEproof}
	\end{proposition}
	
	We can see from \eqref{eq:perm-proj} that $F_{\mathsf{Proj}}(\cdot)$ satisfy the permutation property when the columns and rows in $\mathbf{B}$ can be independently permuted by different matrices, hence the property is called \emph{independent permutation property} in \cite{LSJ_MultiDim_GNN_2022}.
	
	\textbf{Input and output}: To learn the mapping $\mathbf{p}=F_{\mathsf{Proj}}(\mathbf{S,B})$, the inputs are $\mathbf{S}$ and $\mathbf{B}$, and the output is $\hat{\mathbf{p}}\triangleq[\hat{p}_1,\cdots,\hat{p}_K]^T$, where $\hat{p}_k$ is the learned power consumption by $\mathsf{V}_k(\mathbf{r})$.
	
	\textbf{Neural network structure}: The ProjNet can also be designed as a GNN, as to be detailed in section \ref{sec:gnn}. The input-output relationship of the PolicyNet is denoted as,
	\begin{equation}\label{eq:ProjNet}
		\hat{\mathbf{p}}=\mathcal{P}_{\bm\theta_{\sf Proj}}(\mathbf{S,B}),
	\end{equation}
	where $\bm\theta_{\sf Proj}$ denotes all the free parameters  in the ProjNet.
	
	\textbf{Training of the ProjNet}: The ProjNet can be trained with $N_{\mathsf{tr}}$ samples in a supervised manner \footnote{The numbers of samples for training the PolicyNet, the ProjNet and the ValueNet are not necessarily to be the same, but we use $N_{\sf tr}$ to denote the number of training samples for all of the three neural networks for notational simplicity.}. The $i$-th training sample is generated as $\{\mathbf{S}^{[i]}, \mathbf{B}^{[i]}, \mathbf{p}^{[i]}\}$, where $\mathbf{p}^{[i]}=[p_1^{[i]},\cdots, p_k^{[i]}]^T$ is the expected output of power consumption in the $i$-th sample. Since $p_k^{[i]}$ is without closed-form expression, it can be computed numerically, for example, by discretizing the integral region $\mathcal{A}$ into multiple sub-regions and approximate the integral with summation, i.e.,
	\begin{equation}\label{eq:cal-integral}
		p_k^{[i]} = \textstyle\int_{\mathcal{A}} |\mathsf{V}_k^{[i]}(\mathbf{r})|^2 d\mathbf{r} \approx \sum_{m=1}^M |\mathsf{V}_k^{[i]}(\mathbf{r}_m)|^2 \Delta,
	\end{equation}
	where $\mathbf{r}_m$ is the center of the $m$-th sub-region, and $\Delta$ is the area of each sub-region. $\mathsf{V}_k^{[i]}(\mathbf{r})$ is the beamforming of the $k$-th user in the $i$-th sample, which is obtained by $\mathsf{V}_k^{[i]}(\mathbf{r})=\sum_{j=1}^K b_{jk}^{[i]} \mathsf{H}_k^{[i]}(\mathbf{r})$, and $\mathsf{H}_k^{[i]}(\mathbf{r})$ is obtained by substituting $\mathbf{s}_k^{[i]}$ into the channel model in \eqref{eq:chn-model}. The integrals can also be computed numerically with other ways such as Gauss-Legendre quadrature, which can be with lower complexity.
	
	After generating the training samples, the neural network is trained with the following loss function to minimize the mean-squared-error (MSE) between the actual outputs and the expected outputs,
	\begin{equation}\label{eq:loss-proj}
		\mathcal{L}_{\mathsf{Proj}}({\bm\theta}_{\mathsf{Proj}}) = \textstyle\frac{1}{N_{\mathsf{tr}}}\sum_{i\in\mathcal{S}_{\mathsf{Proj}}}\sum_{k=1}^K \big(\hat{p}_k^{[i]} - {p}_k^{[i]}\big)^2,
	\end{equation}
	where $\mathcal{S}_{\mathsf{Proj}}$ denotes the training set for the ProjNet.
	
	\begin{remark}
		To improve the accuracy of approximating integral with summation in \eqref{eq:cal-integral}, the area of sub-region $\Delta$ should be small, which may incur high complexity of computing the integrals. Nonetheless, the process of generating training samples and training the neural network can be conducted offline (e.g., at off-peak time of BSs with sufficient computation resources). Once the neural network is trained, the integrals no longer need to be computed in the online interference phase. 
	\end{remark}
	
	\begin{remark}
		Although the integral in \eqref{eq:cal-integral} is approximated with summation by discretizing the integral region, the learned beamforming is a continuous function, which is different from the optimization-based methods that only optimize beamforming on discontinuous points in the region (e.g., the method ``MIMO'' for comparison in \cite{sanguinetti2022wavenumber}). 
	\end{remark}
	
	After the ProjNet is trained, given the input of $\mathbf{S}$ and $\mathbf{B}$, the ProjNet can output $\hat{\mathbf{p}}$. Then, $\mathbf{b}_k$ can be projected as follows to satisfy the power constraint \eqref{eq:bb-constraint},
	\begin{equation}\label{eq:proj}
		\bar{\mathbf{b}}_k = \mathbf{b}_k\sqrt{P_{\max}}\Big/\sqrt{\textstyle\sum_{j=1}^K \hat{p}_j}. 
	\end{equation}
	
	
	\subsubsection{ValueNet}
	To tackle Challenge 2 that the integrals in \eqref{eq:loss-func} without closed-form expressions hinder training the PolicyNet in an unsupervised manner,
	we design a DNN to learn the integrals in \eqref{eq:bb-object}. The DNN is called ValueNet, because the outputs of the DNN are used to compute the value (i.e., loss function) with the learned policy.
	
	The ValueNet aims to learn the mapping from the finite-dimensional representations of $\mathsf{H}_k(\mathbf{r})$ and $\mathsf{V}_k(\mathbf{r}), k=1,\cdots,K$ to the integrals $g_{kj}\triangleq\int_{\mathcal{A}}\mathsf{H}_k(\mathbf{r})\mathsf{V}_j(\mathbf{r})d\mathbf{r}, k,j=1,\cdots,K$,
	which is denoted as $\mathbf{G}=F_{\mathsf{Value}}(\mathbf{S},\bar{\mathbf{B}})$, where $\bar{\mathbf{B}}=[\bar{\mathbf{b}}_1,\cdots,\bar{\mathbf{b}}_K]$, $\mathbf{G}\in {\mathbb C}^{K\times K}$ is a matrix with the element in the $k$-th row and the $j$-th column being $g_{kj}$. With the same way of analyzing the permutation property in \eqref{eq:perm-policy}, we can obtain the property satisfied by $F_{\mathsf{Value}}(\cdot)$ as follows,
	\begin{proposition}\label{prop4}
		$F_{\mathsf{Value}}(\cdot)$ satisfies the following independent permutation property,
		\begin{equation}\label{eq:perm-value}
			{\bm\Pi}_1^T\mathbf{G}{\bm\Pi}_2=F_{\mathsf{Value}}({\bm\Pi}_1^T\mathbf{S},{\bm\Pi}_1^T\bar{\mathbf{B}}{\bm\Pi}_2).
		\end{equation}
		\begin{IEEEproof}
			The proof is similar to the proof of Proposition \ref{prop3} in Appendix \ref{proof:prop3} and is not provided due to limited space.
		\end{IEEEproof}
	\end{proposition}
	
	\textbf{Input and output}: To learn the function $\mathbf{G}=F_{\mathsf{Value}}(\mathbf{S},\bar{\mathbf{B}})$, the inputs of the ValueNet are $\mathbf{S}$ and $\bar{\mathbf{B}}$, and the output is $\hat{\mathbf{G}}\triangleq[\hat{g}_{kj}]\in{\mathbb C}^{K\times K}$, where $\hat{g}_{kj}$ is the approximated $g_{kj}$.
	
	\textbf{Neural Network Structure}: The neural network can also be designed as a GNN, as to be detailed in section \ref{sec:gnn}. The input-output relationship of the ValueNet is denoted as,
	\begin{equation}\label{eq:ValueNet}
		\hat{\mathbf{G}}=\mathcal{V}_{\bm\theta_{\sf Value}}(\mathbf{S,\bar{B}}).
	\end{equation}
	
	\textbf{Training of the ValueNet}: Similar to the ProjNet, the ValueNet can be trained with $N_{\sf tr}$ samples in a supervised manner. 
	The $i$-th training sample is denoted as $\{\mathbf{S}^{[i]}, \bar{\mathbf{B}}^{[i]}, \mathbf{G}^{[i]}\}$, where the element in the $k$-th row and $j$-th column of $\mathbf{G}^{[i]}$, $g_{kj}^{[i]} = \int_{\mathcal{A}} \mathsf{H}_k^{[i]}(\mathbf{r}){\mathsf{V}}_j^{[i]}(\mathbf{r}) d\mathbf{r}$, can be computed by discretizing $\mathcal{A}$ into multiple sub-regions and approximating the integral with summation, as how $p_k^{[i]}$  is computed in \eqref{eq:cal-integral}. The integrals can also be computed by Gauss-Legendre quadrature.
	
	After generating the training samples, the ValueNet is trained with a loss function to minimize the MSE between the actual outputs and the expected outputs as follows,
	\begin{equation}\label{eq:loss-value}
		\hspace{-2mm}\mathcal{L}_{\mathsf{Value}}({\bm\theta}_{\mathsf{Value}}) =\textstyle \frac{1}{N_{\mathsf{tr}}}\sum_{i\in\mathcal{S}_{\mathsf{Value}}}\sum_{k=1}^K\sum_{j=1}^K \big(\hat{g}_{kj}^{[i]} - {g}_{kj}^{[i]}\big)^2.
	\end{equation}
	where $\mathcal{S}_{\mathsf{Value}}$ denotes the training set for ValueNet.
	
	
	
	\subsection{Training and Inference Procedures} \label{sec:training}
	
	Intuitively, the proposed framework can be trained with the following procedure. The ProjNet and ValueNet are firstly trained with supervised learning. 
	After the two neural networks are trained, 
	given the input of $\mathbf{S}$ and $\bar{\mathbf{B}}$, the matrix of approximated integrals can be obtained as,
	\begin{equation}\label{eq:output}
		\hspace{-2.1mm}\hat{\mathbf{G}}\!=\!\mathcal{V}_{\bm\theta_{\sf Value}^\star}(\mathbf{S,\bar{B}}) 
		\!\overset{(a)}{=}\! \mathcal{V}_{\bm\theta_{\sf Value}^\star}\!\Bigg(\!\mathbf{S},\!\frac{\mathcal{G}_{\bm\theta_p}(\mathbf{S})\sqrt{P_{\max}}}{\sqrt{\!\sum_{j\!=\!1}^K \!\mathcal{P}_{\bm\theta_{\sf Proj}^{\star}}\!\big(\mathbf{S},\mathcal{G}_{\bm\theta_p}(\mathbf{S}) \big)_{\! j}}}\!\Bigg)\!,\!
	\end{equation}
	where $\bm\theta_{\sf Proj}^{\star}$ and $\bm\theta_{\sf Value}^{\star}$ are respectively the trained free parameters of the ProjNet and the ValueNet, $(a)$ comes from substituting \eqref{eq:proj}, \eqref{eq:ProjNet} and \eqref{eq:PolicyNet} into the equation, $\mathcal{P}_{\bm\theta_{\sf Proj}^{\star}}\big(\cdot)_j$ denotes the $j$-th element in the output of the function. We can see that $\hat{\mathbf{G}}$ only depends on $\bm\theta_p$ after the ProjNet and ValueNet are trained.
	Then, the loss function for training the PolicyNet can be computed as follows, which is the negative objective function averaged over all the training samples,
	\begin{equation}\label{eq:loss-policy}
		\mathcal{L}_p(\bm\theta_p)=-\frac{1}{N_{\mathsf{tr}}}\sum_{i=1}^{N_{\mathsf{tr}}} \sum_{k=1}^K\log_2\Bigg(1+\frac{|\mathcal{A}_k|\cdot|\hat{g}_{kk}^{[i]}|^2}{\sum_{j=1,j\neq k}^K |\mathcal{A}_j||\hat{g}_{kj}^{[i]}|^2 + \sigma_k^2}\Bigg).
	\end{equation}
	With the loss function, the gradients can be computed to update $\bm\theta_p$ as,
	{\small
		\begin{equation}\label{eq:gradient}
			\bm\theta_{p} \leftarrow \bm\theta_{p}-\phi\frac{\partial \mathcal{L}_p(\bm\theta_{p})}{\partial \bm\theta_{p}} = \bm\theta_{p} - \frac{\phi}{N_{\sf tr}}\sum_{i=1}^{N_{\sf tr}}\frac{\partial \mathcal{L}_p(\bm\theta_{p})}{\partial \mathbf{G}^{[i]}}\cdot \frac{\partial \mathbf{G}^{[i]}}{\partial \bar{\mathbf{B}}^{[i]}}\cdot \frac{\partial \bar{\mathbf{B}}^{[i]}}{\partial \mathbf{B}^{[i]}}\cdot \frac{\partial \mathbf{B}^{[i]}}{\partial \bm\theta_p},
	\end{equation} }
	
	\noindent where $\frac{\partial \mathbf{G}^{[i]}}{\partial \bar{\mathbf{B}}^{[i]}}$ and $\frac{\partial \bar{\mathbf{B}}^{[i]}}{\partial \mathbf{B}^{[i]}}$ are respectively passed through the trained ValueNet and ProjNet, as shown in Fig. \ref{fig:dl-arch}(b).
	
	In what follows, we refer to this way of firstly training ProjNet and ValueNet and then training PolicyNet as ``Phased Training''. In this way,  the inputs $\mathbf{B}^{[i]}$ and $\bar{\mathbf{B}}^{[i]}$ in the training set of ProjNet and ValueNet are randomly generated, which may not be the same with the outputs of PolicyNet. If there are not enough samples for training ProjNet and ValueNet, the power consumption with \eqref{eq:ProjNet} and the matrix $\hat{\mathbf{G}}$ with \eqref{eq:ValueNet} at the points of PolicyNet's outputs may not be well-approximated. Then, the loss function in \eqref{eq:loss-policy} and its gradient for updating the weights in the PolicyNet are not accurate, which may lead to poor learning performance. 
	
\begin{figure}[!htb]
		\centering
		\includegraphics[width=\linewidth]{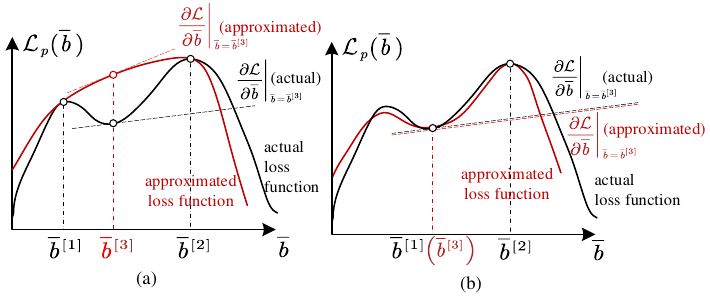}
		\vspace{-5mm}
		\caption{Illustration of actual and approximated loss function with respect to $\mathbf{B}$, where the ValueNet is trained with (a) ``Phased Training'' and (b) ``Alternative Training'' (to be introduced later) methods to approximate the loss function. We consider that $K\!=\!1$ such that $\mathbf{\bar B}$ becomes a scalar $\bar{b}$ and the function can be plotted in a two-dimensional plane.}
		\vspace{-2mm}
		\label{fig:fig-iilus}
	\end{figure}

	In Fig. \ref{fig:fig-iilus} (a), we take training the ValueNet as an example to illustrate this issue, where $\bar{b}^{[1]}$ and $\bar{b}^{[2]}$ are the inputs in two samples for training the ValueNet, and $\bar{b}^{[3]}$ is the output of PolicyNet in one sample. $\bar{b}^{[1]}$ and $\bar{b}^{[2]}$ are randomly generated and are not the same as $\bar{b}^{[3]}$. When the ValueNet is not trained with enough samples, the approximated loss function and its gradient are not accurate at point $\bar{b}^{[3]}$.
	
	To mitigate this issue, we propose another way to train the framework, which is called ``Alternative Training''. Specifically, the PolicyNet, the ProjNet and the ValueNet are alternatively trained, and the training procedure is summarized in Algorithm \ref{algo:training}. In each training epoch, the weights in the PolicyNet are firstly updated. Then, the training samples for the ProjNet are generated, where the input $\mathbf{B}^{[i]}$ is obtained from the output of the ProjNet (step 11), the expected output is obtained from \eqref{eq:cal-integral} (step 12), and the weights in the ProjNet are updated with the generated training samples. After that, the training samples for the ValueNet are generated, where the input $\bar{\mathbf{B}}^{[i]}$ is obtained from projecting the output of the PolicyNet to a feasible region of problem \textbf{P1} with the ProjNet (line 15), and the expected output is generated by computing integrals (line 16). Then, the weights in the ProjNet are updated with the generated training samples.
	
	The advantage of ``Alternative Training'' compared to ``Phased Training'' is as follows. 	With ``Alternative Training'', the training samples for the ProjNet and ValueNet are generated with the output of PolicyNet in each epoch. By training the two DNNs with the generated samples, the power consumption and matrix $\hat{\mathbf{G}}$ at the points of the PolicyNet's outputs can be accurately approximated, such that the loss function and its gradient for updating the weights in the PolicyNet is also accurate for better training the PolicyNet. As illustrated in Fig. \ref{fig:fig-iilus} (b), $\bar{b}^{[1]}$ is generated by the output of PolicyNet $\bar{b}^{[3]}$, i.e., $\bar{b}^{[1]}=\bar{b}^{[3]}$, hence the approximated loss function and its gradient is accurate at this point.


	\begin{algorithm}
		\caption{Alternative Training Procedure}\label{algo:training}
		\small
		\begin{algorithmic}[1]
			\State \textbf{Input}: Training set $\{\mathbf{S}^{[i]}\}_{i=1}^{N_{\mathsf{tr}}}$, number of epochs $N_{\sf epoch}$, learning rate $\phi$.
			\State \textbf{Output}: Optimized free parameters ${\bm \theta}_{\mathsf{p}}^*, {\bm \theta}_{\mathsf{Proj}}^*, {\bm \theta}_{\mathsf{Value}}^*$.
			\State Initialize free parameters ${\bm \theta}_{\mathsf{p}},{\bm \theta}_{\mathsf{Proj}},{\bm \theta}_{\mathsf{Value}}$ and $L_p^\star=0$.
			\For {$m=1:N_{\sf epoch}$}
			
			\hspace{-6mm}\emph{Update free parameters in PolicyNet:}
			
			\State Compute $\hat{\mathbf{G}}^{[i]}$ with input of $\mathbf{S}^{[i]}$ by \eqref{eq:output}, $\forall i$.
			\State Compute the loss function $\mathcal{L}(\bm\theta_{p})$ with \eqref{eq:loss-policy}.
			\If {$\mathcal{L}_p(\bm\theta_{p})<L_p^\star$}
			\State ${\bm \theta}_p^*={\bm \theta}_p, L_p^\star=\mathcal{L}_p(\bm\theta_{p})$;
			\EndIf
			
			\State Compute $\frac{\partial \mathcal{L}(\bm\theta_{p})}{\partial \bm\theta_{p}}$ with \eqref{eq:gradient} and update $\bm\theta_{p} := \bm\theta_{p}-\phi\frac{\partial \mathcal{L}(\bm\theta_{p})}{\partial \bm\theta_{p}}$.
			
			\hspace{-6mm}\emph{Update free parameters in ProjNet:}
			
			\State Compute $\mathbf{B}^{[i]}$ by \eqref{eq:PolicyNet} and compute $\hat{\mathbf{p}}^{[i]}$ by \eqref{eq:ProjNet}, $\forall i$.
			\State Compute expected output $\mathbf{p}^{[i]}$ with \eqref{eq:cal-integral}, $\forall i$.
			\State Compute the loss function $\mathcal{L}(\bm\theta_{\sf Proj})$ with \eqref{eq:loss-proj}.
			
			\State Compute $\frac{\partial \mathcal{L}_{\mathsf{Proj}}(\bm\theta_{\sf Proj})}{\partial \bm\theta_{\sf Proj}}$ and update $\bm\theta_{\mathsf{Proj}}:= \bm\theta_{\sf Proj}-\phi\frac{\partial \mathcal{L}_{\mathsf{Proj}}(\bm\theta_{\sf Proj})}{\partial \bm\theta_{\sf Proj}}$.
			
			\hspace{-6mm}\emph{Update free parameters in ValueNet:}
			
			\State Compute $\mathbf{B}^{[i]}$ by \eqref{eq:PolicyNet}, compute $\hat{\mathbf{p}}^{[i]}$ by \eqref{eq:ProjNet}, compute $\bar{\bf B}^{[i]}$ by
			
			\eqref{eq:proj}, and compute $\hat{\mathbf{G}}^{[i]}$ by \eqref{eq:ValueNet}, $\forall i$.
			\State Compute expected output $\mathbf{G}^{[i]}$ as in \eqref{eq:cal-integral}, $\forall i$.
			\State Compute the loss function $\mathcal{L}(\bm\theta_{\sf Value})$ with \eqref{eq:loss-value}.
			
			\State Compute $\frac{\partial \mathcal{L}_{\mathsf{Value}}(\bm\theta_{\sf Value})}{\partial \bm\theta_{\sf Value}}$ and update $\bm\theta_{\sf Value} := \bm\theta_{\sf Value}-\phi\frac{\partial \mathcal{L}_{\mathsf{Value}}(\bm\theta_{\sf Value})}{\partial \bm\theta_{\sf Value}}$.
			\EndFor
		\end{algorithmic}
	\end{algorithm}
	
	In the inference phase, given the input of $\mathbf{S}$, $\mathbf{B}$ can be outputted by the trained PolicyNet. Then, $\mathbf{B}$ can be projected to $\bar{\mathbf{B}}$ that satisfies \eqref{eq:bb-constraint} with the trained ProjNet, and the learned beamforming can be obtained as $\bar{\mathsf{V}}_k(\mathbf{r})=\sum_{j=1}^K \bar{b}_{jk} \mathsf{H}_k(\mathbf{r})$. The ValueNet is no longer required in this phase, because it is only used to obtain the integrals in the loss function for training the PolicyNet.

	\begin{remark}
		With Proposition \ref{prop1}, the integrals in \eqref{eq:sinr} can be expressed as, 
		\begin{equation}
			\textstyle\int_{\mathcal{A}}\mathsf{H}_k(\mathbf{r}) \mathsf{V}_j(\mathbf{r})d\mathbf{r}\!=\!\textstyle\int_{\mathcal{A}}\mathsf{H}_k(\mathbf{r})\!\sum_{i=1}^K\! b_{ij}\mathsf{H}_i^*(\mathbf{r})d\mathbf{r}\!\triangleq\! \sum_{i=1}^K b_{ij} q_{ki},\notag
		\end{equation}
		where $q_{ki}\triangleq\int_{\mathcal{A}}\mathsf{H}_k(\mathbf{r})\mathsf{H}_i^*(\mathbf{r})d\mathbf{r}$. The integrals in \eqref{eq:bb-constraint} can be expressed as 
		\begin{eqnarray}
			\textstyle\int_{\mathcal{A}}|\mathsf{V}_k(\mathbf{r})|^2 d\mathbf{r}\!\!&\!\!=\!\!&\!\!\textstyle\int_{\mathcal{A}}|\sum_{i=1}^K b_{ik}\mathsf{H}_i^*(\mathbf{r})|^2 d\mathbf{r}\notag\\
			\!\!&\!\!=\!\!&\!\!\textstyle\int_{\mathcal{A}}(\sum_{i=1}^K b_{ik}\mathsf{H}_i^*(\mathbf{r}))(\sum_{j=1}^K b_{jk}^*\mathsf{H}_j(\mathbf{r})) d\mathbf{r}\notag\\
			\!\!&\!\!=\!\!&\!\!\textstyle\sum_{i=1}^K\sum_{j=1}^K b_{ik}b_{jk}^*q_{ji}. \notag
		\end{eqnarray}
		Then, problem \textbf{P1} can be expressed as a parameter optimization problem as,
		\begin{subequations}
			\begin{align}
				\max_{\mathbf{B}}~~ & \sum_{k=1}^K \log_2 \left(1+\frac{|\mathcal{A}_k|\cdot|\mathbf{b}_k^T\mathbf{q}_k|^2}{\sum_{j=1,j\neq k}^K|\mathcal{A}_j|\cdot|\mathbf{b}_j^T\mathbf{q}_k|^2+\sigma_0^2} \right) \notag \\
				{\mathrm{s.t.}} ~~
				&  \textstyle\sum_{k=1}^K\mathsf{Tr}(\mathbf{B}^H\mathbf{QB}) =  P_{\max},\notag
			\end{align}
		\end{subequations}
		where $\mathbf{q}_k=[q_{k1},\cdots,q_{kK}]^T, \mathbf{Q}=[\mathbf{q}_1,\cdots,\mathbf{q}_K]$. After computing the integrals in $\mathbf{Q}$ numerically, the above optimization problem can be solved via iterative algorithms such as weighted-minimum-mean-squared-error (WMMSE) algorithm \cite{WMMSE2011Shi}. When the problem size (say the number of users) is large, computing integrals and running the algorithm may incur high complexity. An alternative way to solve the problem is to resort to the optimal solution structure as follows \cite{Precod_Opt_Structure},
		\begin{equation}\label{eq:opt-structure}
				\mathbf{B}^\star = (\sigma_0^2\mathbf{I}+\mathbf{\Lambda}\mathbf{Q})^{-1}\mathbf{P}^{\frac{1}{2}},
		\end{equation}
		where $\mathbf{\Lambda}={\rm diag}(\lambda_1,\cdots,\lambda_K), \mathbf{P}={\rm diag}(p_1,\cdots,p_K)$ that can be seen as uplink and downlink transmit powers. By learning the powers and the integrals in $\mathbf{Q}$ with DNNs, $\mathbf{B}^\star$ can be obtained with \eqref{eq:opt-structure}. We will investigate this method in future works.
	\end{remark}
	
	\section{\!Learning Beamforming over Graph} \label{sec:gnn}
	In this section, we design the DNNs in DeepCAPA as GNNs that are incorporated with permutation properties. The permutation properties of GNNs depend on their parameter sharing schemes, which further depend on the modeling of graphs (i.e., types of vertices and edges) \cite{GJ_TWC_GNN}.  
	We first show that intuitively modeling the multi-user CAPA system as a heterogeneous graph and applying GNNs to learn the three mappings over the graph causes a permutation property mismatch issue. Then, we show how to model graphs such that the GNNs can satisfy the properties in \eqref{eq:perm-policy}, \eqref{eq:perm-proj} and \eqref{eq:perm-value}.
	
	\subsection{Permutation Property Mismatch Issue}\label{sec:property-mismatch}
\begin{figure}[!htb]
		\centering
		\includegraphics[width=\linewidth]{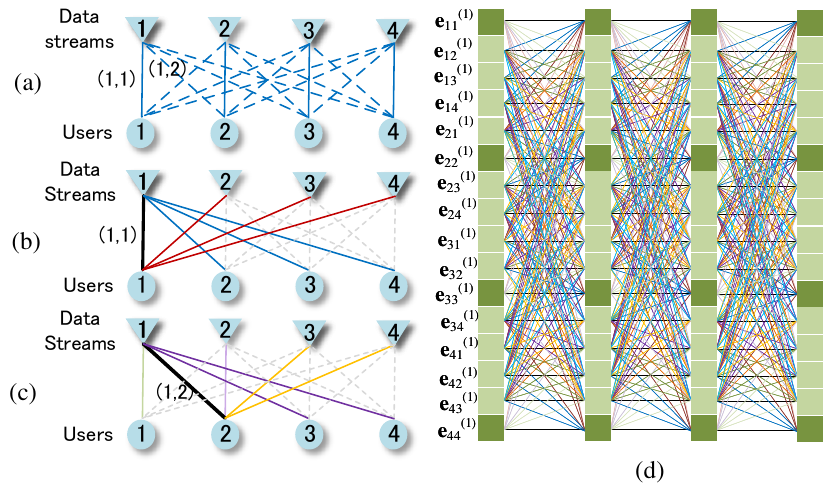}
		\caption{Illustration of (a) a graph $\mathcal{G}_1$ with four data stream vertices and four user vertices, (b) updating the representation of equivalent signal edge $(1,1)$, (c) updating the representation of equivalent interference edge $(1,2)$, and (d) a GNN-$\mathcal{G}_1$ with four layers. In (b), (c) and (d), the same color indicates the same weight matrix, and the dashed line means that the representation of the edge is not aggregated.}
		\vspace{-2mm}
		\label{fig:fig-gnn}
	\end{figure}

	At first glance, the multi-user CAPA system in Fig. \ref{fig:capa} can be modeled as a heterogeneous graph as shown in Fig. \ref{fig:fig-gnn}(a), where there are two types of vertices, i.e., data streams and users, and there are two types of edges, i.e., equivalent signal links and equivalent interference links. The graph is referred to as $\mathcal{G}_1$. Denote the edge from the $k$-th data stream to the $j$-th user as edge $(k,j)$. Two edges are neighbored if they are connected to the same vertex. For example, edge $(k,j)$ is the neighboring edge of edge $(k,k)$ by the $k$-th vertex.
	
	
	There are features and actions on each edge, which are listed below for the three mappings to be learned.
	\begin{itemize}
		\item \textbf{Mapping of PolicyNet (i.e., $\mathbf{B}=F_{\mathsf{P}}(\mathbf{S})$)}:
		\begin{itemize}
			\item \emph{Features}: The feature of edge $(k,k)$ is $\mathbf{s}_k$, and there are no features on edge $(k,j),k\neq j$.
			\item \emph{Actions}: The action of edge $(k,j)$ is $b_{kj}$.
		\end{itemize}
		\item \textbf{Mapping of ProjNet (i.e., $\mathbf{p}=F_{\mathsf{Proj}}(\mathbf{S,B})$)}:
		\begin{itemize}
			\item \emph{Features}: The feature of edge $(k,k)$ is $\{\mathbf{s}_k, b_{kk}\}$, and the feature of edge $(k,j), j\neq k$ is $b_{kj}$.
			\item \emph{Actions}: The action of edge $(k,k)$ is $\hat{p}_k$, and there are no actions on edge $(k,j), j\neq k$.
		\end{itemize}  
		\item \textbf{Mapping of ValueNet (i.e., $\mathbf{G}=F_{\mathsf{Value}}(\mathbf{S},\bar{\mathbf{B}})$)}:
		\begin{itemize}
			\item \emph{Features}: The feature of edge $(k,k)$ is $\{\mathbf{s}_k, \bar{b}_{kk}\}$, and the feature of edge $(k,j),j\neq k$ is $\bar{b}_{kj}$.
			\item \emph{Actions}: The action of  edge $(k,k)$ is $\hat{g}_{kk}$, and the action of edge $(k,j), k\neq j$ is $\hat{g}_{kj}$.
		\end{itemize}  
	\end{itemize}

	Since both the features and the actions are defined on edges, A GNN with $L$ layers can be designed to learn over the graph $\mathcal{G}_1$ (referred to as GNN-$\mathcal{G}_1$), where the representations of the edges are updated in each layer with an \emph{update equation}.  
	
	For updating the representation of each edge in the $(\ell+1)$-th layer (say edge $(k,j)$, which is denoted as $\mathbf{e}_{kj}^{(\ell+1)}$), the representations of neighboring edges in the previous layer are firstly processed by a parameterized \emph{processor} to for information extraction. Then, the extracted information is aggregated with a \emph{pooling function}, and combined with the representation of  edge $(k,j)$ itself with a parameterized \emph{combiner}. 
	
	A parameter sharing scheme is associated with the trainable parameters in the processor by regarding the types of vertices and edges. Specifically, \emph{If some edges are with the same type, and they are connected to a vertex with the same type, then the edges are processed and combined with the same parameters} \cite{GJ_TWC_GNN,PY}. When the processor is linear, the pooling function is summation, and the combiner is a linear function cascaded by an activation function, the equations of updating the representations of equivalent signal edge $(k,k)$ and interference edge $(k,j)$ are as follows \cite{PY},
	\begin{eqnarray}
		\mathbf{e}_{kk}^{(\ell+1)} \!\!&\!\!=\!\!&\!\! \sigma\Big(\mathbf{W}_1\mathbf{e}_{kk}^{(\ell)} +\notag\\
		&& \textstyle\sum_{j=1,j\neq k}^K\mathbf{W}_2\mathbf{e}_{jk}^{(\ell)} + \sum_{j=1,j\neq k}^K\mathbf{W}_3\mathbf{e}_{kj}^{(\ell)} \Big), \notag\\
		\mathbf{e}_{kj}^{(\ell+1)} \!\!&\!\!=\!\!&\!\! \sigma\Big(\textstyle\mathbf{W}_4\mathbf{e}_{kj}^{(\ell)} +  \sum_{i=1,i\neq j,k}^K \mathbf{W}_5\mathbf{e}_{ji}^{(\ell)} + \notag\\
		&&\textstyle\sum_{i=1,i\neq j,k}^K \mathbf{W}_6\mathbf{e}_{ki}^{(\ell)} +\sum_{i=1,i\neq j,k}^K \mathbf{W}_{7}\mathbf{e}_{ij}^{(\ell)} +\notag\\
		&&\textstyle \mathbf{W}_{8}\mathbf{e}_{kk}^{(\ell)} + \mathbf{W}_{9}\mathbf{e}_{jj}^{(\ell)}\Big),\label{eq:upd-gnn-1}
	\end{eqnarray}
	where $\sigma(\cdot)$ denotes the activation function, $\mathbf{W}_1\sim\mathbf{W}_9$ are weight matrices. The parameter sharing scheme for updating the representations of an equivalent signal edge and an equivalent interference edge are respectively illustrated in Fig. \ref{fig:fig-gnn}(b) and (c).

	In the input layer and output layer (i.e., $\ell=1$ and $L$), $\mathbf{e}_{kj}^{(1)}$ and $\mathbf{e}_{kj}^{(L)}$ are respectively the feature and action of edge $(k,j)$. When there are no features or actions for some edges, we set the values of them as zero. For example, for learning the mapping of ProjNet, $\mathbf{e}_{kk}^{(1)}=[\mathbf{s}_k, b_{kk}]$, $\mathbf{e}_{kj}^{(1)}=b_{kj}$, and $\mathbf{e}_{kk}^{(L)}=\hat{p}_k$, $\mathbf{e}_{kj}^{(L)}=\mathbf{0}$. When $1<\ell<L$, the updated representation of each edge is with dimension of $J_{\ell}$, which is a hyper-parameter that can be fine-tuned to achieve the best learning performance.
	
	Fig. \ref{fig:fig-gnn}(d) shows the structure of a GNN with four layers. It can be proved that with this parameter sharing scheme, the input-output relationships of the PolicyNet, ProjNet and ValueNet with this structure respectively satisfy the following permutation properties,
	\begin{eqnarray}
		\text{PolicyNet:~}\bm\Pi^T\mathbf{B}\bm\Pi\!\!&\!\!=\!\!&\!\!\mathcal{G}_{\bm\theta_p}(\bm\Pi^T\mathbf{S}), \label{eq:perm-policy-gnn}\\
		\text{ProjNet:~}\bm\Pi^T\hat{\mathbf{p}}\!\!&\!\!=\!\!&\!\!\mathcal{P}_{\bm\theta_{\sf Proj}}(\bm\Pi^T\mathbf{S},\bm\Pi^T\mathbf{B}\bm\Pi), \label{eq:perm-proj-gnn}\\
		\text{ValueNet:~}\bm\Pi^T\hat{\mathbf{G}}\bm\Pi\!\!&\!\!=\!\!&\!\!\mathcal{V}_{\bm\theta_{\sf Value}}(\bm\Pi^T\mathbf{S},\bm\Pi^T\bar{\mathbf{B}}\bm\Pi). \label{eq:perm-value-gnn}
	\end{eqnarray}
	
	We can see that all of the three properties are dependent permutation properties, where the permutations of the rows or columns of the variables depend on the permutations of users. The property in \eqref{eq:perm-policy-gnn} matches the property of the beamforming policy in \eqref{eq:perm-policy}, but the properties of the ProjNet and the ValueNet do not match the independent properties in \eqref{eq:perm-proj} and \eqref{eq:perm-value}.
	
	This property mismatch issue causes learning inefficiency. Specifically, the permutation property is a prior knowledge that can restrict the hypothesis space of functions learnable by a DNN, such that the training complexity (including the number of samples and time of training) for finding the optimal mappings can be reduced. For learning the mappings $F_{\mathsf{Proj}}(\cdot)$ and $F_{\mathsf{Value}}(\cdot)$, the prior knowledge introduced into the GNN-$\mathcal{G}_1$ is not sufficient, such that the input-output relationships of the GNNs only satisfy the permutation property when the columns and rows are permuted by the same matrix $\bm\Pi$ instead by arbitrary matrices $\bm\Pi_1$ and $\bm\Pi_2$. Hence, the hypothesis space of GNN-$\mathcal{G}_1$ is larger than DNNs with permutation properties satisfying \eqref{eq:perm-proj} and \eqref{eq:perm-value}, and the training complexity of the (say the training time) for finding the optimal mappings is also higher.
	
	In what follows, we show how to model a graph and design GNNs to learn the mappings $\mathbf{p}=F_{\mathsf{Proj}}(\mathbf{S,B})$ and $\mathbf{G}=F_{\mathsf{Value}}(\mathbf{S},\bar{\mathbf{B}})$ with permutation properties respectively satisfying \eqref{eq:perm-proj} and \eqref{eq:perm-value}. Then, we explain why the GNNs designed in this section cannot satisfy the properties in \eqref{eq:perm-proj} and \eqref{eq:perm-value}.
	
	\subsection{GNNs for Learning $F_{\mathsf{Proj}}(\cdot)$ and $F_{\mathsf{Value}}(\cdot)$}
	
\begin{figure}[!htb]
		\centering
		\includegraphics[width=\linewidth]{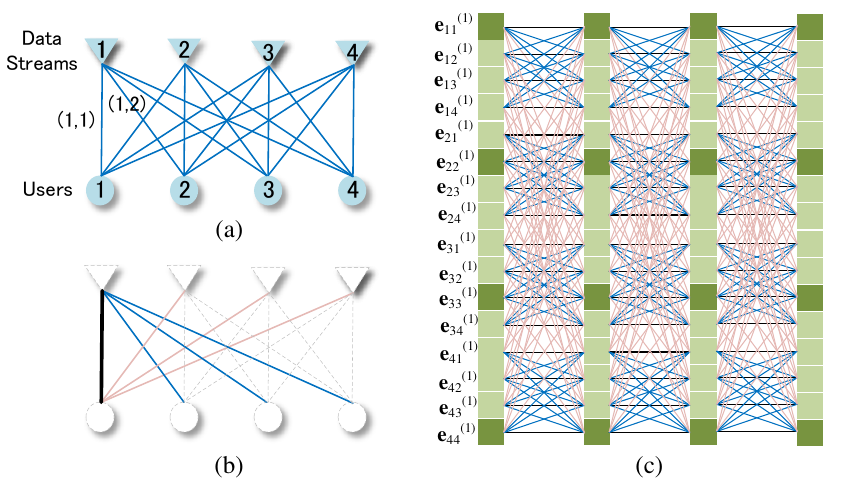}
		\caption{Illustration of (a) a graph $\mathcal{G}_2$ with four data streams and four users, (b) updating the representation of edge $(1,1)$ and (d) a GNN-$\mathcal{G}_2$ with four layers. In (b) and (c), the same color indicates the same weight matrix, and the dashed vertices and edges mean that the representations of the vertices or edges are not aggregated.}
		\vspace{-2mm}
		\label{fig:fig-gnn-2}
	\end{figure}

	Since the two mappings satisfy independent permutation properties, they can be learned over a heterogeneous graph shown in Fig. \ref{fig:fig-gnn-2}(a), where there are two types of vertices (i.e., data streams and users), and there is only one type of edges (instead of two types of edges as in graph $\mathcal{G}_1$) between the two types of vertices \cite{LSJ_MultiDim_GNN_2022}. The graph is referred to as $\mathcal{G}_2$.
	
	For learning $F_{\mathsf{Proj}}(\cdot)$ with the ProjNet, the feature of the $k$-th data stream vertex and the edge $(k,j)$ are respectively $\mathbf{s}_k$ and $b_{kj}$, and the action of the $j$-th user vertex is $\hat{p}_j$. For learning $F_{\mathsf{Value}}(\cdot)$ with the ValueNet, the feature of the $k$-th data stream vertex and the edge $(k,j)$ are respectively $\mathbf{s}_k$ and $\bar{b}_{kj}$, and the action of the edge $(k,j)$ is $\hat{g}_{kj}$. Then, the permutations $\bm\Pi_1$ and $\bm\Pi_2$ in \eqref{eq:perm-proj} and \eqref{eq:perm-value} respectively correspond to independently permuting data stream vertices and user vertices. 
	
	A GNN can be applied to learn over the  graph $\mathcal{G}_2$ (referred to as GNN-$\mathcal{G}_2$), which contains $L$ layers, and the representations of all the edges are updated in each layer by an update equation, also with a processor, a pooling function and a combiner. Parameter sharing is associated with the GNN by regarding to the types of vertices and edges, as have stated above. The representation of the edge $(k,j)$ is updated as follows,
	\begin{equation}\label{eq:upd-gnn-2}
		\mathbf{e}_{kj}^{(\ell+1)} = \textstyle\sigma\Big(\mathbf{W}_1\mathbf{e}_{kj}^{(\ell)}+\sum_{i=1,i\neq j}^K\mathbf{W}_2 \mathbf{e}_{ki}^{(\ell)} + \sum_{i=1,i\neq k}^K\mathbf{W}_3 \mathbf{e}_{ij}^{(\ell)}\Big),
	\end{equation}
	where $\mathbf{W}_1$ is the weight matrix in the combiner, $\mathbf{W}_2$ and $\mathbf{W}_3$ are the weight matrices in the processors for extracting information of neighboring edges by the $k$-th data stream vertex and the $j$-th user vertex.
	
	In the input and output layers of the ProjNet, $\mathbf{e}_{kj}^{(1)}=[\mathbf{s}_k,b_{kj}]$ and $\mathbf{e}_{kj}^{(L)}=\hat{p}_j$. In the input and output layers of the ValueNet, $\mathbf{e}_{kj}^{(1)}=[\mathbf{s}_k,\bar{b}_{kj}]$ and $\mathbf{e}_{kj}^{(L)}=\hat{g}_{kj}$. Fig. \ref{fig:fig-gnn-2}(c) shows the structure of the GNN with four layers. It can be proved that the input-output relationships of the ProjNet and the ValueNet with this structure match the permutation properties in \eqref{eq:perm-proj} and \eqref{eq:perm-value}, i.e., 
	\begin{eqnarray}
		\text{ProjNet:~}\bm\Pi_2^T\hat{\mathbf{p}}\!\!&\!\!=\!\!&\!\!\mathcal{P}_{\bm\theta_{\sf Proj}}(\bm\Pi_1^T\mathbf{S},\bm\Pi_1^T\mathbf{B}\bm\Pi_2), \label{eq:perm-proj-gnn1}\\
		\text{ValueNet:~}\bm\Pi_1^T\hat{\mathbf{G}}\bm\Pi_2\!\!&\!\!=\!\!&\!\!\mathcal{V}_{\bm\theta_{\sf Value}}(\bm\Pi_1^T\mathbf{S},\bm\Pi_1^T\bar{\mathbf{B}}\bm\Pi_2). \label{eq:perm-value-gnn1}
	\end{eqnarray}
	
	\subsection{Why the Properties Mismatch?}
	
	We can see from Fig. \ref{fig:capa} and the analysis of the permutation property of the beamforming policy in section \ref{sec:policynet} that the dependent permutation property of the policy comes from the corresponding relationship between the data streams and users (i.e., the $k$-th data stream is transmitted the $k$-th user), which further comes from the data rate formula in \eqref{eq:bb-object}. Specifically, only when the data streams and the users are permuted in the same way, the corresponding relationship between the data streams and users are preserved, and hence the data rate in \eqref{eq:bb-object} and the policy obtained from problem \textbf{P1} keep unchanged. Hence, by modeling the CAPA system as a graph as shown in Fig. \ref{fig:fig-gnn}(a) with two different types of edges, the input-output relationship of the GNN is only equivariant to the same permutations of data stream vertices and user vertices, which matches the dependent permutation property of the beamforming policy. 
	
	By contrast, the mappings $\mathbf{p}=F_{\mathsf{Proj}}(\mathbf{S,B})$ and $\mathbf{G}=F_{\mathsf{Value}}(\mathbf{S},\bar{\mathbf{B}})$ do not depend on the data rate formula, and hence do not depend on the serving relationship between data streams and users. Therefore, the two mappings should be learned over the graph in Fig. \ref{fig:fig-gnn-2}(a) with only one type of edges, such that there are no corresponding relationships between data streams and users, and the two types of vertices can be permuted independently. If the mappings are still learned over the graph in Fig. \ref{fig:fig-gnn}(a) that preserves the serving relationship after permutations of vertices, then the learned mappings satisfy dependent permutation properties that are mismatched with the properties of $F_{\mathsf{Proj}}(\cdot)$ and $F_{\mathsf{Value}}(\cdot)$.
	
	We can see from the analyses that even though all of the PolicyNet, ProjNet and ValueNet serve for learning the beamforming policy in the CAPA system, they should not be simply designed as the same GNN structure that learns over the same graph, which causes property mismatch issue. A promising approach to avoid this issue is to first analyze the permutation property satisfied by each mapping to be learned, and then model graphs according to the permutation properties as in \cite{LSJ_MultiDim_GNN_2022} and use GNNs to learn over the graphs, or directly design parameter sharing scheme in DNNs to satisfy the properties as in \cite{PEandParamShare2017}.
	
	\section{Numerical Results}\label{sec:results}
	In this section, we first provide numerical results to validate the impact of property mismatch issue and the superiority of training DeepCAPA with the  ``Alternative Training'' way. Then, we validate that the trained DeepCAPA can achieve the performance upper bound of SE in the SPD system as the number of antennas in a fixed-sized area approaches infinity. Finally, we show the learning performance of DeepCAPA under different scenarios.  
	
	Consider that a BS deployed with a planar CAPA transmits to $K$ users. The aperture size of the CAPA is $|\mathcal{A}|=0.25~{\rm m}^2$ unless otherwise specified. Each user's position can be expressed as a Cartesian coordinate $(x,y,z)$. The users are uniformly deployed in a squared region with coordinates in the range of $(\pm1,\pm1,30)$ m. The normal vector of the CAPA and the wavelength are respectively set as $\mathbf{e}_r=[0,1,0]$ and $\lambda=0.0107$ m \cite{zhao2024continuous}. Without loss of generality, we set $|\mathcal{A}_k|=A=\frac{\lambda^2}{4\pi}$ and $\sigma_k=\sigma_0$, which are the same for all the users. Then, the signal-to-noise-plus-interference ratio (SINR) in \eqref{eq:sinr} can be written as follows,
	\begin{eqnarray}
		\gamma_k \!\!&\!\!=\!\!&\!\! \frac{A\cdot|\int_{\mathcal{A}} \mathsf{H}_k(\mathbf{r}) \mathsf{V}_k(\mathbf{r}) d\mathbf{r}|^2}{\sum_{j=1,j\neq k}^K A\cdot |\int_{\mathcal{A}} \mathsf{H}_k(\mathbf{r}) \mathsf{V}_j(\mathbf{r}) d\mathbf{r}|^2 + \sigma_0^2} \notag\\
		\!\!&\!\!=\!\!&\!\! \frac{\zeta\cdot|\int_{\mathcal{A}} \mathsf{H}_k'(\mathbf{r}) \mathsf{V}_k'(\mathbf{r}) d\mathbf{r}|^2}{\sum_{j=1,j\neq k}^K \zeta\cdot |\int_{\mathcal{A}} \mathsf{H}_k'(\mathbf{r}) \mathsf{V}_j'(\mathbf{r}) d\mathbf{r}|^2 + 1},\notag
	\end{eqnarray}
	where $\zeta=\frac{A}{\sigma_0^2}\frac{k_0^2\eta^2}{4\pi}P_{\max}$, $\mathsf{H}_k'(\mathbf{r}) = \frac{2\pi^{\frac{1}{2}}}{k_0\eta}\mathsf{H}_k(\mathbf{r})$ and $\mathsf{V}_k'(\mathbf{r})=\frac{1}{\sqrt{P_{\max}}}\mathsf{V}_k(\mathbf{r})$. In the simulations, we control the signal-to-noise ratio (SNR) by changing $\zeta$.
	
	
	
	We compare the performance of the learning-based methods with three numerical algorithms as follows,
	\begin{itemize}
		\item \textbf{Match-filtering (MF)}: The conjugate of channel function of the $k$-th user normalized by the transmit power is adopted as the beamformer, i.e.,
		\begin{equation}
			\mathsf{V}_k(\mathbf{r}) = \frac{\mathsf{H}^*_k(\mathbf{r})}{\sqrt{\int_{\mathcal{A}}|\mathsf{H}^*_k(\mathbf{r})|^2d\mathbf{r}}}\cdot \sqrt{p_k}, \notag
		\end{equation}
		where the integral is computed by summation over the discretized integral region as in \eqref{eq:cal-integral}, and $p_k$ is the transmit power for the $k$-th user that is optimized from the WMMSE algorithm \cite{WMMSE2011Shi}. This method can maximize the desired signal of each user but cannot well-suppress multi-user interference. 
		\item \textbf{WMMSE}: We discretize the CAPA into $M$ squared regions, and the channel response in each region is approximated by the channel response at the center of the region. Then, the original problem is reduced to a beamforming optimization problem in the SPD system, where the beamforming vectors for the users can be optimized with the WMMSE algorithm \cite{WMMSE2011Shi}. The optimal beamforming vector is also a linear combination of channel vectors.  By multiplying the linear combination vectors with $\mathsf{H}_1(\mathbf{r}),\cdots, \mathsf{H}_K(\mathbf{r})$, we can obtain an approximated solution of problem \textbf{P1}.
		\item \textbf{Fourier}: This is the method proposed in \cite{zhang2023pattern}, where the continuous current distribution is approximated with finite Fourier series, such that problem \textbf{P1} is reduced to a parameter optimization problem to optimize the projection lengths on the Fourier basis functions.
	\end{itemize}
	
	\subsection{Impact of Property Mismatch Issue}
\begin{figure}[!htb]
		\centering
		\vspace{-3mm}
		\includegraphics[width=.9\linewidth]{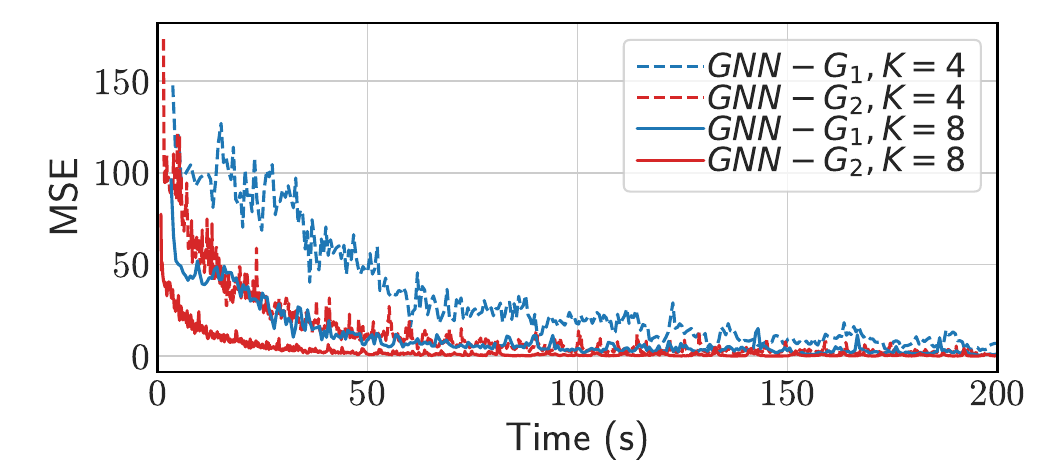}
		\vspace{-2mm}
		\caption{Convergence of GNN-$\mathcal{G}_1$ and GNN-$\mathcal{G}_2$ in the training phase.}
		\label{fig:mmse-time}
		\vspace{-2mm}
	\end{figure}

	We first show the impact of property mismatch issue on learning efficiency, by taking training the ValueNet as an example. Fig. \ref{fig:mmse-time} shows how the value of MSE in \eqref{eq:loss-value} changes with training time, where the ValueNet is respectively set as a GNN-$\mathcal{G}_1$ and a GNN-$\mathcal{G}_2$. It can be seen that the GNN-$\mathcal{G}_2$ converges faster than the GNN-$\mathcal{G}_1$, i.e., less training time is required to achieve an expected performance. This is because the property satisfied by the GNN-$\mathcal{G}_1$ does not match the property satisfied by \eqref{eq:ValueNet}, such that the number of trainable parameters is much higher than the GNN-$\mathcal{G}_2$ with matched property. This can be seen from \eqref{eq:upd-gnn-1} and \eqref{eq:upd-gnn-2}, i.e., nine and three weight matrices need to be trained in each layer of GNN-$\mathcal{G}_1$  and GNN-$\mathcal{G}_2$, respectively.
	\emph{This simulation result validates the impact of property mismatch issue and the benefit of incorporating proper prior knowledge to efficient learning}.
	
	We can also see from the figure that the MSE achieved when $K=8$ is much lower than the MSE achieved when $K=4$. The reason was explained in \cite{GJ-RGNN}. To be brief, since the weight matrices are the same for all the users, a sample generated in a scenario with $K$ users can be seen as $K$ equivalent samples for training the weight matrices. Then, the weight matrices can be more efficiently trained with samples generated in scenarios with more users.
	
	In what follows, we refer to the framework with PolicyNet being GNN-$\mathcal{G}_1$, ProjNet and ValueNet being GNN-$\mathcal{G}_2$ as DeepCAPA (GNN), where proper permutation properties are incorporated with the GNNs. To further show the impact of incorporating with the permutation properties on the learning performance, we compare the performance of DeepCAPA (GNN) with another framework, namely DeepCAPA (FNN), which is also with the architecture in Fig. \ref{fig:dl-arch}(b) but the PolicyNet, ProjNet and ValueNet are FNNs without incorporating with any properties. The fine-tuned hyper-parameters of the two frameworks are given in Table \ref{table:hyper-params}.
	
	\begin{table}[!htb]
		\centering
		\caption{Hyper-parameters of DeepCAPA frameworks, $K=4$.}\label{table:hyper-params}
		\vspace{-2mm}
		\footnotesize
		\setlength\tabcolsep{0.5pt}
		\begin{tabular}{cc|c|c|c|c}
			\hline\hline
			\multicolumn{2}{c|}{\textbf{Framework}}                                                                                         & \textbf{\begin{tabular}[c]{@{}c@{}}Num. of \\      hidden \\ layers\end{tabular}} & \textbf{\begin{tabular}[c]{@{}c@{}}Num. of \\      neurons in each\\      hidden layer\end{tabular}} & \textbf{\begin{tabular}[c]{@{}c@{}}Learning\\      rate\end{tabular}} & \textbf{\begin{tabular}[c]{@{}c@{}}Activation  \\ function\\      of output layer\end{tabular}} \\ \hline
			\multicolumn{1}{c|}{\multirow{3}{*}{\textbf{\begin{tabular}[c]{@{}c@{}}DeepCAPA\\      (GNN)\end{tabular}}}} & \textbf{PolicyNet} & 5                                                                              & {[}16, 32, 64, 32, 16{]}                                                                             & 0.001                                                                 & Tanh                                                                                          \\ \cline{2-6} 
			\multicolumn{1}{c|}{}                                                                                      & \textbf{ProjNet}   & 4                                                                              & {[}4, 8, 8, 4{]}                                                                                     & 0.0001                                                                & Relu                                                                                          \\ \cline{2-6} 
			\multicolumn{1}{c|}{}                                                                                      & \textbf{ValueNet}  & 6                                                                              & \begin{tabular}[c]{@{}c@{}}{[}16,   32, 64, \\      64, 32, 16{]}\end{tabular}                       & 0.001                                                                 & Linear                                                                                        \\ \hline
			\multicolumn{1}{c|}{\multirow{3}{*}{\textbf{\begin{tabular}[c]{@{}c@{}}DeepCAPA\\      (FNN)\end{tabular}}}} & \textbf{PolicyNet} & 5                                                                              & \begin{tabular}[c]{@{}c@{}}{[}256,   512, 1024, \\       512, 256{]}\end{tabular}                                                                     & 0.001                                                                 & Tanh                                                                                          \\ \cline{2-6} 
			\multicolumn{1}{c|}{}                                                                                      & \textbf{ProjNet}   & 4                                                                              & {[}64, 128, 128, 64{]}                                                                               & 0.0001                                                                & Relu                                                                                          \\ \cline{2-6} 
			\multicolumn{1}{c|}{}                                                                                      & \textbf{ValueNet}  & 6                                                                              & \begin{tabular}[c]{@{}c@{}}{[}256,   512, 1024, \\      1024, 512, 256{]}\end{tabular}               & 0.001                                                                 & Linear                                                                                        \\ \hline\hline
		\end{tabular}
	\vspace{-5mm}
	\end{table}
	
	\subsection{Performance of Training DeepCAPA with Different Ways}
\begin{figure}[!htb]
		\centering
		\vspace{-4mm}
		\includegraphics[width=\linewidth]{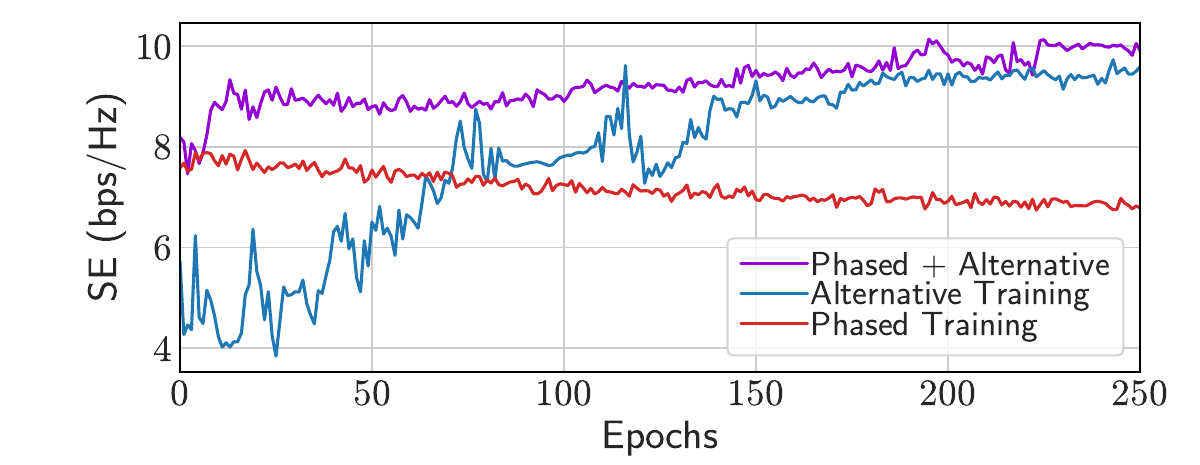}
		\vspace{-5mm}
		\caption{Convergence in the training phase of DeepCAPA.}
		\label{fig:perf-epoch}
		\vspace{-2mm}
	\end{figure}

	In Fig. \ref{fig:perf-epoch}, we show how the SE achieved by the PolicyNet changes with the epochs during the training phase, where the PolicyNet is trained with the two ways in section \ref{sec:training}. 
	We can see that the SE achieved with the ``Phased Training'' method degrades with training epochs, and the achieved SE after convergence is much lower than the ``Alternative Training'' method. The results validate the superiority of the ``Alternative Training'' method compared to the ``Phased Training'' method, and the reason was explained in section \ref{sec:training}.
	
	We can also see from the figure that compared to the ``Phased Training'' method, the SE achieved by ``Alternative Training'' in the first 100 epochs is very low. This is because the ProjNet and ValueNet are not trained in advance, which causes a ``cold start'' issue that the loss function and gradients in the earlier epochs are not well-estimated. To resolve this issue, we combine the ``Phased Training'' and ``Alternative Training'' methods. Specifically, the ProjNet and ValueNet are firstly trained before training PolicyNet, and then fine-tuned by alternative training with the PolicyNet. We can see from the figure that this method (labeled with ``Phased + Alternative'') can achieve high SE both in earlier epochs and after convergence. In what follows, we use this method to train the proposed framework.
	
	\subsection{SE Upper Bound of WMMSE Algorithm when $M\to\infty$}
\begin{figure}[!htb]
		\centering
		\includegraphics[width=0.75\linewidth]{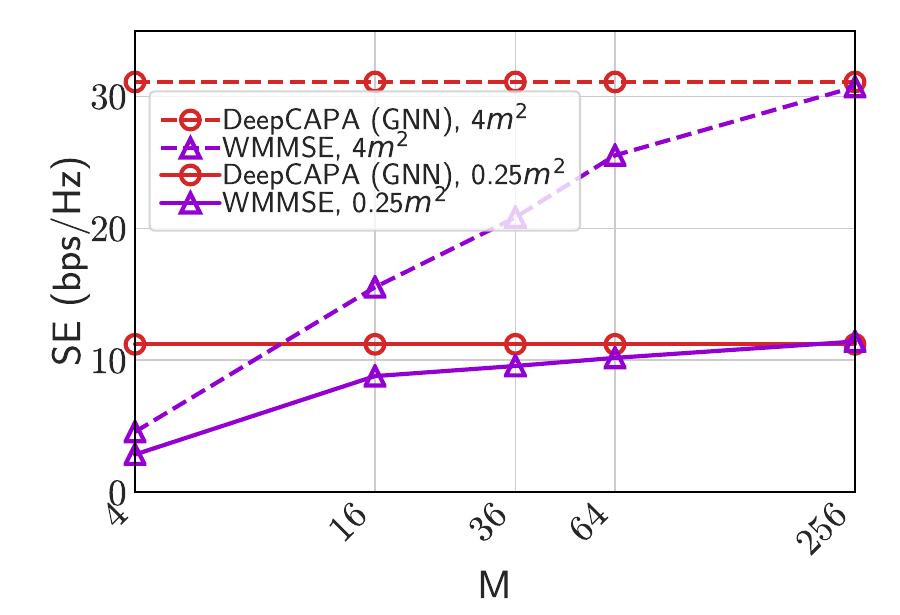}
		\vspace{-2mm}
		\caption{SE versus $M$, $K=4$, SNR=50 dB. }
		\label{fig:perf-wm}
		\vspace{-5mm}
	\end{figure}

	In Fig. \ref{fig:perf-wm}, we show how the SE achieved by the WMMSE algorithm changes with $M$, where the aperture size is fixed. We also provide the performance of the DeepCAPA (GNN) trained with 50000 samples. It can be seen that the SE grows slower with $M$ and is close to but does not exceed the SE achieved by the DeepCAPA (GNN), which validates that the learned beamforming can achieve the performance upper bound of optimizing beamforming in the SPD system when $M\to\infty$. We can also see that the performance of the WMMSE algorithm converges faster to the upper bound when the aperture size is small, because in this case the channel vector in the discretized SPD system can well-approximate the channel function in the CAPA system even with a small value of $M$.
	
	\subsection{Learning Performance in Different Scenarios}
\begin{figure*}[!htb]
		\centering
		\begin{minipage}[t]{0.3\linewidth}
			\centering
			\includegraphics[width=\textwidth]{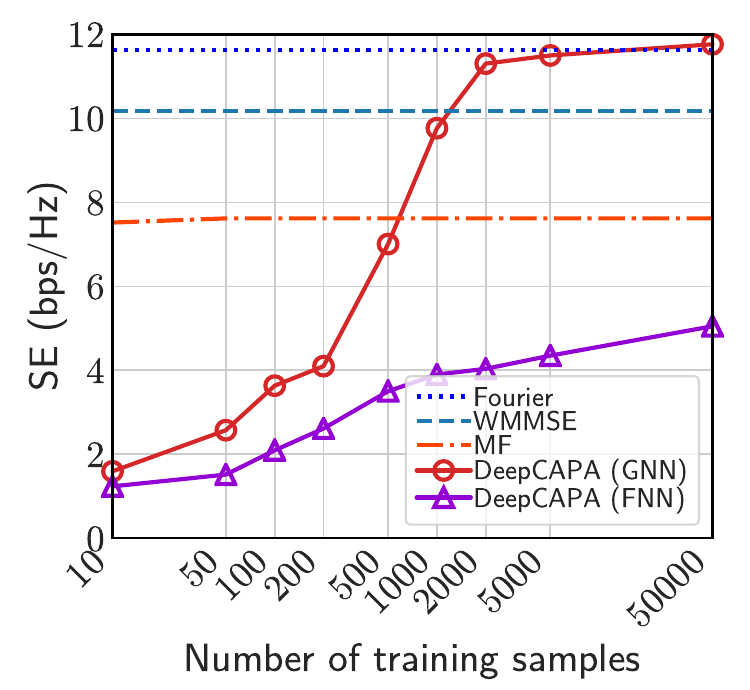}
			\caption{SE versus $N_{\mathsf{tr}}$, $K=4, M=36$, SNR=50 dB.}
			\label{fig:perf-ntr}
		\end{minipage}\hspace{4mm}
		\begin{minipage}[t]{0.3\linewidth}
			\centering
			\includegraphics[width=\textwidth]{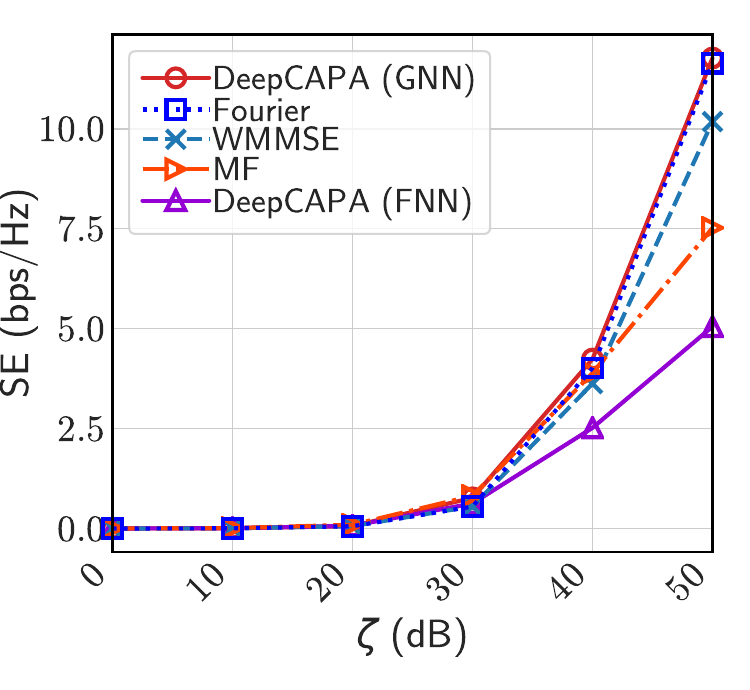}
			\caption{SE versus SNR, $K=4, M=36$, $N_{\mathsf{tr}}=50000$. }
			\label{fig:perf-snr}
		\end{minipage}\hspace{4mm}
		\begin{minipage}[t]{0.3\linewidth}
			\centering
			\includegraphics[width=\textwidth]{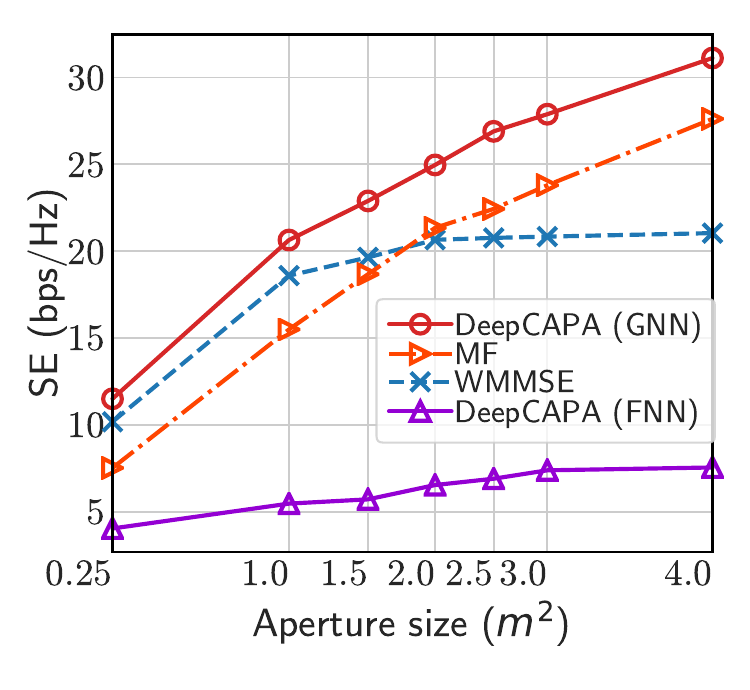}
			\caption{SE versus aperture size, $K=4, M=36$, SNR=50 dB, $N_{\mathsf{tr}}=50000$. }
			\label{fig:perf-apsize}
		\end{minipage}\hspace{3mm}
		
		\vspace{-4mm}
	\end{figure*}

	Fig. \ref{fig:perf-ntr} shows how the learning performance changes with the number of training samples, where the number of discretized squared regions (i.e., $M$) for the WMMSE algorithm is not large for affordable computational complexity. In this simulation and in the following simulations, we assume that all the users are allocated with equal power, such that we can better show the effectiveness of each method on finding null space of channel functions (i.e., $\int_{\mathcal{A}}\mathsf{V}_k(\mathbf{r})\mathsf{H}_j(\mathbf{r})d\mathbf{r}=0$) for suppressing inter-user interference.  It can be seen from the figure that the DeepCAPA (GNN) can achieve higher SE than MF and WMMSE with more than 2000 training samples. This is because MF cannot well-suppress multi-user interference as stated above, and the discretization in WMMSE incurs performance loss. We can also see that the DeepCAPA (GNN) performs close to or slightly better than the Fourier-based method with more than 5000 training samples. 
	The SE achieved by the DeepCAPA (GNN) is much higher than the SE achieved by the DeepCAPA (FNN). The performance gain comes from leveraging the permutation properties. This result also indicates that the number of samples required for the DeepCAPA (GNN) to achieve an expected performance (i.e., sample complexity) is much lower than the DeepCAPA (FNN).

	In Fig. \ref{fig:perf-snr}, we show the SE under different values of SNR. 
	We can see from the figure that the SE achieved by DeepCAPA (GNN) is higher than the SE achieved by the numerical algorithms and DeepCAPA (FNN), and the reason has been explained above. We can also see that the performance gap between DeepCAPA (GNN) and MF is smaller in the low SNR region. This is because such a region corresponds to a noise-limited scenario where the noise power is much higher than the interference power, In this scenario, the gain brought by suppressing inter-user interference is marginal, and MF is already near-optimal. The DeepCAPA (GNN) can achieve a slightly higher SE than the Fourier-based method, with much lower computational complexity, as to be shown later.
	
	Fig. \ref{fig:perf-apsize} shows the achieved SE in scenarios with different aperture sizes. It can be seen that the performance gap between the DeepCAPA (GNN) and WMMSE is larger when the aperture size is large. The reason is as follows. Recall that the aperture is discretized into $M$ regions. When the number of discretized regions (i.e., $M$) does not change, the discretized antenna array cannot well-approximate the CAPA when the aperture size is large. Hence, the performance degradation brought by discretization is large. The performance of the Fourier-based method is not provided, because when $|\mathcal{A}|$ is larger than $0.25~{\rm m}^2$, the running complexity of the method is too high and causes an out-of-memory issue.
	
	
	We next compare the average inference time of the DeepCAPA (GNN) with the average running time of the WMMSE algorithm and the Fourier-based method over all the test samples. The results are obtained on a computer with one 14-core Intel i9-9940X CPU, one Nvidia RTX 2080Ti GPU, and 64 GB memory. For the WMMSE algorithm, we set the number of discretized regions as $M=256$ such that the algorithm can achieve a close-to-upper-bound performance, as shown in Fig. \ref{fig:perf-wm}. As can be seen from Table \ref{table:time}, the inference time of ``DeepCAPA (GNN)'' is much lower than the running time of the WMMSE algorithm and the Fourier-based methods. The running time of the Fourier-based method when $|\mathcal{A}|$ is larger than $0.25~{\rm m}^2$ is still not provided, also because the complexity is too high and causes an out-of-memory issue. The running time of MF is also not provided, because it cannot achieve high enough SE as demonstrated in Fig. \ref{fig:perf-ntr}$\sim$Fig. \ref{fig:perf-apsize}.
	
	\begin{table}[]
		\centering
		\caption{Time consumption (in seconds) in scenarios with different aperture sizes, $K=4$}\label{table:time}
		\vspace{-2mm}
		\footnotesize
		\setlength\tabcolsep{5pt}
		\begin{tabular}{llllllll}
			\hline\hline
			\multicolumn{1}{c|}{$|\mathcal{A}|$   (m$^2$)}  & \multicolumn{1}{c|}{0.25}  & \multicolumn{1}{c|}{1}    & \multicolumn{1}{c|}{1.5}   & \multicolumn{1}{c|}{2}    & \multicolumn{1}{c|}{2.5}  & \multicolumn{1}{c|}{3}    & \multicolumn{1}{c}{4}    \\ \hline
			\multicolumn{1}{c|}{DeepCAPA (GNN)} & \multicolumn{1}{c|}{0.02}  & \multicolumn{1}{c|}{0.02} & \multicolumn{1}{c|}{0.02}  & \multicolumn{1}{c|}{0.03} & \multicolumn{1}{c|}{0.03} & \multicolumn{1}{c|}{0.03} & \multicolumn{1}{c}{0.03} \\ \hline
			\multicolumn{1}{c|}{WMMSE}          & \multicolumn{1}{c|}{43.5}  & \multicolumn{1}{c|}{47.6} & \multicolumn{1}{c|}{43.67} & \multicolumn{1}{c|}{50.7} & \multicolumn{1}{c|}{45.3} & \multicolumn{1}{c|}{47.5} & \multicolumn{1}{c}{48.3} \\ \hline
			\multicolumn{1}{c|}{Fourier}        & \multicolumn{1}{c|}{308.5} & \multicolumn{6}{c}{---}                                                                                                                                               \\ \hline\hline
		\end{tabular}
		\vspace{-5mm}
	\end{table}
	
	
	\section{Conclusions}\label{sec:conclusions}
	In this paper, we proposed a DeepCAPA framework to learn beamforming policies in CAPA systems. The framework includes a PolicyNet that maps the channel responses to the beamforming solutions, a ProjNet and a ValueNet that respectively learns the integrals in the power constraint and the objective function of the optimization problem. The finite-dimensional representations of the functions of channel and beamforming were found such that they can be inputted into and outputted from the DNNs. The ProjNet and ValueNet were trained such that the gradients of the loss function can be back-propagated for training the PolicyNet. The three neural networks were designed as GNNs to leverage the permutation properties of the mappings to be learned. Numerical results demonstrated the performance of DeepCAPA in achieving high SE under different system settings with short running time. The proposed framework can also be applied to learn other policies in CAPA systems, such as EE-maximal beamforming, power allocation and user scheduling.
	
	%
	\vspace{-1mm}
	\begin{appendices}
		\numberwithin{equation}{section}
		\section{Proof of Proposition \ref{prop1}}\label{proof:prop1}
		We prove this by proving its converse-negative proposition: if $\exists k$ such that the optimal $\mathsf{V}_k(\mathbf{r})$ cannot be expressed as $\mathsf{V}_k(\mathbf{r})=\sum_{j=1}^K b_{jk} \mathsf{H}_j^*(\mathbf{r})$, then there exists a better solution that can achieve higher SE while satisfying  \eqref{eq:bb-constraint}.
		
		If $\exists k$ such that the optimal  $\mathsf{V}_k(\mathbf{r})$ is not in the subspace spanned by $\mathsf{H}_1^*(\mathbf{r}),\cdots,\mathsf{H}_K^*(\mathbf{r})$ (called conjugate channel subspace in the sequel), then $\mathsf{V}_k(\mathbf{r})$ can be expressed as,
		\begin{equation}\label{eq:perp}
			\mathsf{V}_k(\mathbf{r})=\mathsf{V}_{k//}(\mathbf{r})+\mathsf{V}_{k\perp}(\mathbf{r}),
		\end{equation}
		where $\mathsf{V}_{k//}(\mathbf{r})$ is in the conjugate channel subspace, while $\mathsf{V}_{k\perp}(\mathbf{r})$ is not, i.e., $\int_{\mathcal{A}}\mathsf{V}_{k\perp}(\mathbf{r})\mathsf{H}_j(\mathbf{r})d\mathbf{r}=0 , \forall j$, and
		\begin{equation}\label{eq:perp1}
			\textstyle\int_{\mathcal{A}}\mathsf{V}_{k\perp}(\mathbf{r})\mathsf{V}_{k//}(\mathbf{r})d\mathbf{r}=0.
		\end{equation}
		In this case, the SE can be expressed as,
		\begin{equation}
			R_0=\sum_{j=1}^K \log_2\Bigg(1+\frac{|\mathcal{A}_j|\!\cdot\!|\int_{\mathcal{A}} \mathsf{H}_j(\mathbf{r}) \mathsf{V}_j'(\mathbf{r}) d\mathbf{r}|^2}{\sum_{i=1,i\neq j}^K\! |\mathcal{A}_i|\!\cdot\!|\int_{\mathcal{A}} \mathsf{H}_j(\mathbf{r}) \mathsf{V}_i'(\mathbf{r}) d\mathbf{r}|^2 \!+\! \sigma_i^2}\Bigg),\notag
		\end{equation}
		where 
		\begin{equation}\label{eq:perp2}
			\mathsf{V}_j'(\mathbf{r})=\left\{
			\begin{aligned}
				\mathsf{V}_j(\mathbf{r}), & ~~j \neq k, \\
				\mathsf{V}_{j//}(\mathbf{r}), &~~ j=k.
			\end{aligned}
			\right.
		\end{equation}
		
		With \eqref{eq:perp} and \eqref{eq:perp1}, we have $\int_{\mathcal{A}}|\mathsf{V}_k(\mathbf{r})|^2d\mathbf{r}=\int_{\mathcal{A}}|\mathsf{V}_{k//}|^2(\mathbf{r})d\mathbf{r}+\int_{\mathcal{A}}\mathsf|{V}_{k\perp}(\mathbf{r})|^2d\mathbf{r}+2\int_{\mathcal{A}}|\mathsf{V}_{k//}(\mathbf{r})\mathsf{V}_{k\perp}(\mathbf{r})|d\mathbf{r}=\int_{\mathcal{A}}|\mathsf{V}_{k//}(\mathbf{r})|^2 d\mathbf{r}+\int_{\mathcal{A}}|\mathsf{V}_{k\perp}(\mathbf{r})|^2 d\mathbf{r}$, hence $\int_{\mathcal{A}}|\mathsf{V}_k(\mathbf{r})|^2d\mathbf{r}>\int_{\mathcal{A}}|\mathsf{V}_{k//}(\mathbf{r})|^2 d\mathbf{r}$. 
		
		We now construct a solution of \textbf{P1} as
		$\bar{\mathsf{V}}_j(\mathbf{r}) = C\mathsf{V}_j'(\mathbf{r})$ with
		\begin{equation}
			C = \sqrt{\frac{P_{\max}}{\sum_{j=1}^K \int_{\mathcal{A}}|\mathsf{V}_j'(\mathbf{r})|^2 d\mathbf{r}}} = \sqrt{\frac{\sum_{j=1}^K \int_{\mathcal{A}}|\mathsf{V}_j(\mathbf{r})|^2 d\mathbf{r}}{\sum_{j=1}^K \int_{\mathcal{A}}|\mathsf{V}_j'(\mathbf{r})|^2 d\mathbf{r}}}.\notag
		\end{equation}
		This is a feasible solution of \textbf{P1}, because it satisfies constraint \eqref{eq:bb-constraint}. From \eqref{eq:perp2} and $\int_{\mathcal{A}}|\mathsf{V}_k(\mathbf{r})|^2d\mathbf{r}>\int_{\mathcal{A}}|\mathsf{V}_{k//}(\mathbf{r})|^2 d\mathbf{r}$, we can obtain that $C>1$. Then, the SE achieved by the constructed solution is
		
		{\small
			\begin{eqnarray}
				R_1\!\!\!&\!\!\!=\!\!\!&\!\!\!\sum_{j=1}^K \log_2\Bigg(\!1+\frac{|\mathcal{A}_j|\cdot|\int_{\mathcal{A}} \mathsf{H}_j(\mathbf{r}) \mathsf{V}_j'(\mathbf{r})\cdot C d\mathbf{r}|^2 }{\sum_{i=1,i\neq j}^K |\mathcal{A}_i|\cdot|\int_{\mathcal{A}} \mathsf{H}_j(\mathbf{r}) \mathsf{V}_i'(\mathbf{r})\cdot C d\mathbf{r}|^2 + \sigma_i^2}\!\Bigg) \notag\\
				\!\!\!&\!\!\!=\!\!\!&\!\!\!\sum_{j=1}^K\! \log_2\!\Bigg(\!1\!+\!\frac{|\mathcal{A}_j|\cdot|\int_{\mathcal{A}} \mathsf{H}_j(\mathbf{r}) \mathsf{V}_j'(\mathbf{r}) d\mathbf{r}|^2 }{\sum_{i=1,i\neq j}^K |\mathcal{A}_i|\!\cdot\!|\int_{\mathcal{A}} \mathsf{H}_j(\mathbf{r}) \mathsf{V}_i'(\mathbf{r}) d\mathbf{r}|^2 \!+\! \frac{\sigma_i^2}{C^2}}\!\Bigg)\!> \! R_0. \notag
			\end{eqnarray}
		}
		
		\noindent This indicates that the feasible solution we constructed can achieve a higher SE than $\mathsf{V}_1(\mathbf{r}),\cdots,\mathsf{V}_K(\mathbf{r})$, hence $\{\mathsf{V}_1(\mathbf{r}),\cdots,\mathsf{V}_K(\mathbf{r})\}$ is not the optimal solution.
		
		\section{Proof of Proposition \ref{prop2}}\label{proof:prop2}
		We prove this proposition by proving that the objective function and constraint in problem \textbf{P1} keep unchanged after permuting $\mathbf{S}$ and $\mathbf{B}$ to $\mathbf{\Pi}^T\mathbf{S}$ and $\mathbf{\Pi}^T\mathbf{B}\mathbf{\Pi}$. By proving this,  given $\mathbf{\Pi}^T\mathbf{S}$,  the optimal solution to problem \textbf{P1} is $\mathbf{\Pi}^T\mathbf{B}\mathbf{\Pi}$, and hence the property in \eqref{eq:perm-policy} holds.
		
		The objective function of problem \textbf{P1} can be written as,
		
		{\small
			\begin{eqnarray}
				&&\hspace{-7mm}O = \sum_{k=1}^K \log_2 \left(1+\frac{A|\int_{\mathcal{A}} \mathsf{H}_k(\mathbf{r}) \mathsf{V}_k(\mathbf{r}) d\mathbf{r}|^2}{\sum_{j=1,j\neq k}^K A|\int_{\mathcal{A}} \mathsf{H}_k(\mathbf{r}) \mathsf{V}_j(\mathbf{r}) d\mathbf{r}|^2 + \sigma_k^2} \right)\notag\\
				&&\hspace{-7mm}=\sum_{k=1}^K \log_2 \!\left(\!1\!+\!\frac{A|\int_{\mathcal{A}} \sum_{i=1}^K b_{ik}|\mathsf{H}(\mathbf{r},\mathbf{s}_k)|^2 d\mathbf{r}|^2}{\sum\limits_{j=1,j\neq k}^K A\Big|\int_{\mathcal{A}} \sum\limits_{i=1}^K b_{ij}\mathsf{H}(\mathbf{r},\mathbf{s}_k) \mathsf{H}^*(\mathbf{r},\mathbf{s}_i) d\mathbf{r}\Big|^2 \!+\! \sigma_k^2} \!\right). \notag
		\end{eqnarray}}
		
		\noindent When $\mathbf{B}$ is permuted to $\bm\Pi^T\mathbf{B}\bm\Pi$ and $\mathbf{S}$ is permuted to $\bm\Pi^T\mathbf{S}$, $b_{ij}$ is permuted to $b_{\pi(i)\pi(j)}$, and $\mathbf{s}_k$ is permuted to  $\mathbf{s}_{\pi(k)}$. Then, the objective function becomes,
		
		{\small
			\begin{eqnarray}
				&&\hspace{-7mm}\sum_{k=1}^K \log_2 \!\!\left(\!1\!+\!\frac{A|\int_{\mathcal{A}} \sum_{i=1}^K b_{\pi(i)\pi(k)}|\mathsf{H}(\mathbf{r},\mathbf{s}_{\pi(k)})|^2 d\mathbf{r}|^2}{\!\!\sum\limits_{j=1,j\neq k}^K \!\! \! A\Big|\int_{\!\mathcal{A}} \sum\limits_{i=1}^K b_{\pi(i)\pi(j)}\mathsf{H}(\mathbf{r},\mathbf{s}_{\pi(k)}) \mathsf{H}^*(\mathbf{r},\mathbf{s}_{\pi(i)}) d\mathbf{r}\Big|^2 \!\!\!+\! \sigma_k^2} \!\!\right)\notag\\
				&&\hspace{-7mm}\overset{(a)}{=}\sum_{k=1}^K \log_2 \!\left(\!1\!+\!\frac{A|\int_{\mathcal{A}} \sum_{i=1}^K b_{i\pi(k)}|\mathsf{H}(\mathbf{r},\mathbf{s}_{\pi(k)})|^2 d\mathbf{r}|^2}{\sum\limits_{j=1,j\neq k}^K \!\!A\Big|\int_{\mathcal{A}} \sum\limits_{i=1}^K b_{ij}\mathsf{H}(\mathbf{r},\mathbf{s}_{\pi(k)}) \mathsf{H}^*(\mathbf{r},\mathbf{s}_{i}) d\mathbf{r}\Big|^2 \!+\! \sigma_k^2}\! \right)\notag\\
				&&\hspace{-7mm}\overset{(b)}{=}\sum_{k=1}^K \log_2 \!\left(\!1\!+\!\frac{A|\int_{\mathcal{A}} \sum_{i=1}^K b_{ik}|\mathsf{H}(\mathbf{r},\mathbf{s}_{k})|^2 d\mathbf{r}|^2}{\!\!\sum\limits_{j=1,j\neq k}^K \!\! A\Big|\int_{\mathcal{A}} \sum\limits_{i=1}^K b_{ij}\mathsf{H}(\mathbf{r},\mathbf{s}_{k}) \mathsf{H}^*(\mathbf{r},\mathbf{s}_{i}) d\mathbf{r}\Big|^2 \!\!+\! \sigma_k^2} \!\!\right) \!=\! O.\notag
		\end{eqnarray}}
		
		\noindent where $(a)$ and $(b)$ come from the fact that changing the order of elements in a summation term does not change the summation result.
		
		With the same way of proving the objective function does not change after permutations, it is not hard to prove that the constraint in \eqref{eq:bb-constraint} also does not change after permutations.
		
		\vspace{-1mm}
		\section{Proof of Proposition \ref{prop3}}\label{proof:prop3}\vspace{-1mm}
		The power consumption of the $k$-th user can be written as,
		\begin{eqnarray}
			p_k \!\!&\!\!=\!\!&\!\! \textstyle\int_{\mathcal{A}} |\mathsf{V}_k(\mathbf{r})|^2 d\mathbf{r} 
			= \textstyle\int_{\mathcal{A}} |\sum_{j=1}^K b_{jk}\mathsf{H}_j^*(\mathbf{r})|^2 d\mathbf{r} \notag\\
			\!\!&\!\!=\!\!&\!\! \textstyle\int_{\mathcal{A}} |\sum_{j=1}^K b_{jk}\mathsf{H}^*(\mathbf{r},\mathbf{s}_j)|^2 d\mathbf{r}.\notag
		\end{eqnarray}
		When the rows and columns of $\mathbf{B}$ are permuted to $\bm\Pi_1^T\mathbf{B}\bm\Pi_2$, and the rows of $\mathbf{S}$ are permuted to $\bm\Pi_1^T\mathbf{S}$, $b_{jk}$ is permuted to $b_{\pi_1(j)\pi_2(k)}$, and $\mathbf{s}_j$ is permuted to $\mathbf{s}_{\pi_1(j)}$. Then, $p_k$ becomes,
		\begin{eqnarray}
			p_k' \!\!&\!\!=\!\!&\!\! \textstyle\int_{\mathcal{A}} |\sum_{j=1}^K b_{\pi_1(j)\pi_2(k)}\mathsf{H}^*(\mathbf{r},\mathbf{s}_{\pi_1(j)})|^2 d\mathbf{r}\notag\\
			\!\!&\!\!=\!\!&\!\! \textstyle\int_{\mathcal{A}} |\sum_{j=1}^K b_{j\pi_2(k)}\mathsf{H}^*(\mathbf{r},\mathbf{s}_{j})|^2 d\mathbf{r} = p_{\pi_2(k)}.\notag
		\end{eqnarray}
		This indicates that $\mathbf{p}$ becomes $\bm\Pi_2^T\mathbf{p}$ after permutations. Hence, the permutation property in \eqref{eq:perm-proj} holds.
		
	\end{appendices}

\vspace{-1mm}
	\bibliography{IEEEabrv,GJ}

\end{document}